 \documentclass[final,5p,times,twocolumn]{elsarticle}

\usepackage[T1]{fontenc} % Use modern font encodings

\usepackage{amssymb}

\usepackage{textcomp}
\usepackage{xspace}
\usepackage{xcolor}
\usepackage{txfonts}
\usepackage{booktabs}

\usepackage{hyperref}

\journal{ACS Earth and Space Chemistry}

\begin{document}

\begin{frontmatter}

%%%%%%%%%%%%%%%%%%%%%%%%%%%%%%%%%%%%%%%%%%%%%%%%%%%%%%%%%%%%%%%%%%%%%
%% The document title should be given as usual. Some journals require
%% a running title from the author: this should be supplied as an
%% optional argument to \title.
%%%%%%%%%%%%%%%%%%%%%%%%%%%%%%%%%%%%%%%%%%%%%%%%%%%%%%%%%%%%%%%%%%%%%
\title{Rotational Spectroscopy as a Tool to Study Vibration-Rotation Interaction: 
   Investigations of $^{13}$CH$_3$CN and CH$_3$$^{13}$CN up to $\varv _8 = 2$ 
   and a Search for $\varv _8 = 2$ Transitions toward Sagittarius B2(N)}

\author[Koeln]{Holger S.~P. M\"uller\corref{cor}} 
\ead{hspm@ph1.uni-koeln.de} 
\cortext[cor]{Corresponding author.} 
\author[Bonn]{Arnaud Belloche} 
\author[Koeln]{Frank Lewen} 
\author[Koeln]{Stephan Schlemmer}

\address[Koeln]{I.~Physikalisches Institut, Universit{\"a}t zu K{\"o}ln, 
  Z{\"u}lpicher Str. 77, 50937 K{\"o}ln, Germany}
\address[Bonn]{Max-Planck-Institut f{\"u}r Radioastronomie, Auf dem H{\"u}gel 69, 
  53121 Bonn, Germany}

%%%%%%%%%%%%%%%%%%%%%%%%%%%%%%%%%%%%%%%%%%%%%%%%%%%%%%%%%%%%%%%%%%%%%
%% Some journals require a list of abbreviations or keywords to be
%% supplied. These should be set up here, and will be printed after
%% the title and author information, if needed.
%%%%%%%%%%%%%%%%%%%%%%%%%%%%%%%%%%%%%%%%%%%%%%%%%%%%%%%%%%%%%%%%%%%%%
%% \abbreviations{IR}

\begin{abstract}
Methyl cyanide, CH$_3$CN, is present in diverse regions in space, 
in particular in the warm parts of star-forming regions where it is a common molecule. 
Rotational transitions of $^{13}$CH$_3$CN and CH$_3$$^{13}$CN in their 
$\varv _8 = 1$ lowest excited vibrational states ($E_{\rm vib} \approx 520$~K) 
are quite prominent in Sagittarius B2(N). 
In order to be able to search for transitions of the next higher vibrational state $\varv _8 = 2$, 
we recorded spectra of samples enriched in $^{13}$CH$_3$CN and CH$_3$$^{13}$CN up to $\varv _8 = 2$ 
in the 35 to 1091~GHz region and reinvestigated existing spectra of CH$_3$CN 
in its natural isotopic composition between 1085 and 1200~GHz. 
Perturbations caused by near-degeneracies in $K = 4$ of $\varv _8 = 2^0$ 
and $K = 2$ of $\varv _8 = 2^{-2}$ yielded accurate information on the 
energy spacing of 22.93 and 21.79~cm$^{-1}$ between the $l$-components of 
$^{13}$CH$_3$CN and CH$_3$$^{13}$CN, respectively. 
Fermi-type interaction between $K = 13$ and 14 of $\varv _8 = 1^{-1}$ and $\varv _8 = 2^{+2}$ 
probe the energy differences between the two states of both isotopomers. 
In addition, a $\Delta K \pm2$, $\Delta l \mp1$ interaction between 
the ground vibrational state of $^{13}$CH$_3$CN and $\varv _8 = 1^{+1}$ provides information on their energy spacing. 
Furthermore, we obtained improved or extended ground state rotational transition frequencies of 
$^{13}$CH$_3$$^{13}$CN and extensive data for $^{13}$CH$_3$C$^{15}$N and CH$_3$$^{13}$C$^{15}$N. 
Finally, we report the results of our search for transitions of 
$^{13}$CH$_3$CN and CH$_3$$^{13}$CN in their $\varv _8 = 2$ states toward Sagittarius B2(N).
\end{abstract}


\end{frontmatter}


%%%%%%%%%%%%%%%%%%%%%%%%%%%%%%%%%%%%%%%%%%%%%%%%%%%%%%%%%%%%%%%%%%%%%
%%%%%%%%%%%%%%%%%%%%%%%%%%%%%%%%%%%%%%%%%%%%%%%%%%%%%%%%%%%%%%%%%%%%%
\section{INTRODUCTION}
\label{intro}
%%%%%%%%%%%%%%%%%%%%%%%%%%%%%%%%%%%%%%%%%%%%%%%%%%%%%%%%%%%%%%%%%%%%%
%%%%%%%%%%%%%%%%%%%%%%%%%%%%%%%%%%%%%%%%%%%%%%%%%%%%%%%%%%%%%%%%%%%%%

Methyl cyanide, also known as acrylonitrile or cyanomethane, is important in astrochemistry and in astrophysics. 
It was detected in Sagittarius (Sgr) A and B more than 50 years ago as one of the first molecules 
observed by radioastronomical means \cite{MeCN_det_T_1971} and the 17th molecule detected in space \cite{astrochymist}.
Already this first report on CH$_3$CN in space pointed out its potential as a probe into the excitation temperature 
in these sources because transitions with the same overall rotational quantum number $J$ but different $K$ occur 
in fairly narrow frequency ranges, but sample very different energies. 
Methyl cyanide was not only found in high-mass star forming regions, but also in the warm and dense 
parts of molecular clouds surrounding low-mass protostars \cite{MeCN_IRAS_16293_2003}, 
in cold, dark molecular clouds \cite{MeCN-TMC-1_1983}, in the envelopes of carbon-rich late-type stars \cite{MeCN-CSE_1984}, 
in external galaxies \cite{MeCN_MeCCH_extragal}, in disks around young protostars \cite{MeCN_PP-disk_2015}, 
and in translucent molecular clouds \cite{translucent-toward-SgrB2_2019}.

The molecule is also a trace constituent in Earth's atmosphere, 
predominantly caused by biomass burning \cite{biomass_burning_MLS_2004,biomass_burning_2011}. 
Concerning the Solar system, it was also found in comets, such as Kohoutek \cite{MeCN-Kohoutek_1974}, 
and in the atmosphere of Titan \cite{MeCN-Titan_1993}.

Methyl cyanide was not only identified in space in its ground vibrational state, but was also detected in its 
$\varv _8 = 1$ ($E_{\rm{vib}} = 525$~K) \cite{CH3CN_v8eq1_det_1983}, $\varv _8 = 2$ \cite{OrionKL-ALMA_CH3CNv8eq2_2012}, 
$\varv _4 = 1$ ($E_{\rm{vib}} = 1324$~K) \cite{SgrB2-survey_2013}, and $\varv _4 = \varv _8 = 1$ \cite{MeCN_up2v4eq1_etc_2021} 
excited vibrational states, albeit in the last state only tentatively.

Many minor isotopic species of CH$_3$CN were detected in space (here and in the following, unlabeled atoms refer to 
$^1$H, $^{12}$C, and $^{14}$N), which are CH$_3$$^{13}$CN \cite{CH3CN_w_CH313CN_SgrB2_1983}, $^{13}$CH$_3$CN \cite{Orion-A_survey_1985}, 
CH$_2$DCN \cite{det_CH2DCN_1992}, CH$_3$C$^{15}$N \cite{SgrB2_survey_1998}, 
and even $^{13}$CH$_3$$^{13}$CN \cite{EMoCA_2016} and CHD$_2$CN shortly thereafter \cite{CHD2CN_det_2018}. 
In the case of $^{13}$CH$_3$CN and CH$_3$$^{13}$CN, also transitions in $\varv _8 = 1$ were identified \cite{SgrB2-survey_2013}, 
making transitions in $\varv _8 = 2$ a next logical target.

M{\"u}ller et al. \cite{MeCN_rot_2009} provided a comprehensive account on the ground state rotational spectra of CH$_3$CN, 
all singly substituted isotopologs, as well as $^{13}$CH$_3$$^{13}$CN in natural isotopic composition. 
In a later study of rotational and rovibrational spectra of CH$_3$CN up to $\varv _8 = 2$ \cite{MeCN_v8le2_2015}, 
a resonance between $\varv = 0$ and $\varv _8 = 1$ was analyzed in detail for the first time 
and revealed Fermi-type interactions ($\Delta K = 0$, $\Delta l = \pm3$) between $\varv _8 = 1$ and 2 as well as between $\varv _8 = 2$ and 3 
besides numerous other resonances, of which several were not reported previously. 
A more recent investigation included rotational and rovibrational spectra of CH$_3$CN in $\varv _4 = 1$ 
into this analysis with additions or improvements in parts of the lower state data 
and with the identification of further resonances, some identified newly \cite{MeCN_up2v4eq1_etc_2021}.

M{\"u}ller et al. \cite{MeCN_isos_v8_rot_2016} presented extensive analyses of $^{13}$CH$_3$CN, CH$_3$$^{13}$CN, and CH$_3$C$^{15}$N 
in their $\varv _8 = 1$ states using a sample in natural isotopic composition. 
Even though the accessed $K$ levels did not reach high enough to cover the region of stronger interactions between $\varv _8 = 1$ and 2, 
it was still necessary to account for these interactions by estimating $\varv _8 = 2$ parameters 
as well as some interaction parameters based on those of the main isotopic species. 
These reports are testament to methyl cyanide also being of great interest for basic science, in particular 
spectroscopy, and as a test case for quantum chemistry.

As the $\varv _8 = 1$ lines of $^{13}$CH$_3$CN and CH$_3$$^{13}$CN in natural isotopic composition 
were already too weak prior to reaching $K$ level experiencing the strongest $\varv _8 = 1$/2 perturbations, 
our present investigation is based on new spectral recordings between 35 and 1091~GHz employing samples 
enriched in $^{13}$CH$_3$CN and CH$_3$$^{13}$CN, respectively. 
We did study these isotopomers not only in their $\varv _8 = 2$ states, but also in $\varv = 0$  and $\varv _8 = 1$. 
As added boni, we extended the data set of $^{13}$CH$_3$$^{13}$CN from the previous study \cite{MeCN_rot_2009} 
and obtained extensive line lists for $^{13}$CH$_3$C$^{15}$N and CH$_3$$^{13}$C$^{15}$N, for which only limited data 
up to 72~GHz were available thus far \cite{13C15N-MeCN-rot_1989}.

The remainder of the paper is organized as follows. 
Information on the measurements is given in \autoref{exptl}, 
\autoref{intro-spec} provides clues on the rotational and vibrational spectroscopy of CH$_3$CN isotopologs relevant to this study, 
and \autoref{spcat-spfit} reveals how the rotational spectra were calculated and fit. 
The results of our laboratory measurements are described in \autoref{lab-results}, 
\autoref{lab-discussion} discusses these results, 
\autoref{astro} presents the astronomical results, 
and conclusions are given in \autoref{conclusion}.

%%%%%%%%%%%%%%%%%%%%%%%%%%%%%%%%%%%%%%%%%%%%%%%%%%%%%%%%%%%%%%%%%%%%%
%%%%%%%%%%%%%%%%%%%%%%%%%%%%%%%%%%%%%%%%%%%%%%%%%%%%%%%%%%%%%%%%%%%%%
\section{EXPERIMENTAL DETAILS}
\label{exptl}
%%%%%%%%%%%%%%%%%%%%%%%%%%%%%%%%%%%%%%%%%%%%%%%%%%%%%%%%%%%%%%%%%%%%%
%%%%%%%%%%%%%%%%%%%%%%%%%%%%%%%%%%%%%%%%%%%%%%%%%%%%%%%%%%%%%%%%%%%%%

Measurements at the Universit{\"a}t zu K{\"o}ln were carried out between 35 and 1091~GHz applying three slightly different spectrometer setups. 
All utilize glass cells of different lengths and 100~mm inner diameter kept at room temperature. 
All new measurements were carried out employing methyl cyanide samples enriched in $^{13}$CH$_3$CN or CH$_3$$^{13}$CN 
(Sigma-Aldrich Chemie GmbH) with 99\% $^{13}$C at the respective position; there was no indication of $^{13}$C enrichment at the other C atom. 
Sample pressures were in the range of 0.5 to 1.0~Pa in most cases; pressures of 0.1~Pa or less were applied for measurements of 
the low $K$ ($\le 12$) transitions in the ground vibrational states, whereas pressures of 2.5~Pa or slightly higher were 
chosen for very weak lines. 
No unexpected or unusually high safety hazards were encountered in the course of our investigations. 
In order to minimize any risk, it is strongly recommended to follow safety precautions listed in the safety data sheets 
and use small quantities, about 1~g at most, preferably less, at a time.

An Agilent E8257D synthesizer at fundamental frequencies was used for measurements between 35 and 56~GHz, 
while this synthesizer together with a VDI frequency tripler was employed between 71 and 111~GHz. 
The measurements were done in two connected 7~m long glass cells in a single path arrangement. 
This setup is a modification of the one employed in ref.~\citenum{n-BuCN_rot_2012}.

Various active and passive frequency multipliers from the VDI starter kit driven by Rohde \& Schwarz SMF~100A or Keysight E8257D synthesizers were used for higher frequencies. 
A 5~m long double path cell served for measurements in the 160 to 252~GHz region. 
The spectrometer was described in detail earlier \cite{OSSO_rot_2015}. 
The glass cells at these lower frequencies were equipped with Teflon windows and Schottky-diode detectors.

Measurements between 355 and 1091~GHz were carried out in a 5~m long single path cell sealed with HDPE windows. 
A closed-cycle liquid He cooled bolometer (QMC Instruments Ltd) served as detector. Frequency modulation was applied throughout; 
the demodulation at $2f$ causes the lines to appear similar to a second derivative of a Gaussian. 
Further information on this spectrometer system is available elsewhere \cite{MeSH_rot_2012}.

Part of the spectral recordings were broader scans, frequently covering about one vibrational state of one isotopolog 
with integration times adjusted to achieve very good signal-to-noise ratios (S/N) for fairly weak lines. 
Very weak lines were recorded individually with integration times adjusted to reach at a least good S/N in each case. 
In addition, some of the stronger lines were recorded in single line scans.

We also inspected broad frequency scans of CH$_3$CN in natural isotopic composition covering 1083$-$1200~GHz and 
taken much earlier \cite{MeCN_rot_2009} with the JPL cascaded multiplier spectrometer \cite{JPL_multiplier_spectrometer_2005}. 
These scans have especially good S/N among the JPL scans and are mostly higher in frequency than recordings taken in Cologne. 
A multiplier chain source is passed through a one to two-meter path length flow cell 
and is detected by a silicon bolometer cooled to near 1.7~K for these recordings. 
The cell is filled with a steady flow of reagent grade acetonitrile. 
Pressure and modulation are optimized to yield good S/N with narrow lineshapes. 
The S/N was optimized for a higher $K$ ground state transition ($K = 12$), such that the lower $K$ transitions 
exhibit saturated line profiles. This way, a better dynamic range was obtained for lines of the 
rare isotopologs and for highly excited vibrational satellites.

%%%%%%%%%%%%%%%%%%%%%%%%%%%%%%%%%%%%%%%%%%%%%%%%%%%%%%%%%%%%%%%%%%%%%
\section{SPECTROSCOPIC PROPERTIES OF \texorpdfstring{CH$_3$CN}{CH3CN}}
%%%%%%%%%%%%%%%%%%%%%%%%%%%%%%%%%%%%%%%%%%%%%%%%%%%%%%%%%%%%%%%%%%%%%
\label{intro-spec}
%%%%%%%%%%%%%%%%%%%%%%%%%%%%%%%%%%%%%%%%%%%%%%%%%%%%%%%%%%%%%%%%%%%%%
%%%%%  Figure 1  %%%%%%%%%%%%%%%%%%%%%%%%%%%%%%%%%%%%%%%%%%%%%%%%%%%%
%%%%%%%%%%%%%%%%%%%%%%%%%%%%%%%%%%%%%%%%%%%%%%%%%%%%%%%%%%%%%%%%%%%%%

 \begin{figure}
 \begin{center}
  \includegraphics[width=5.0cm]{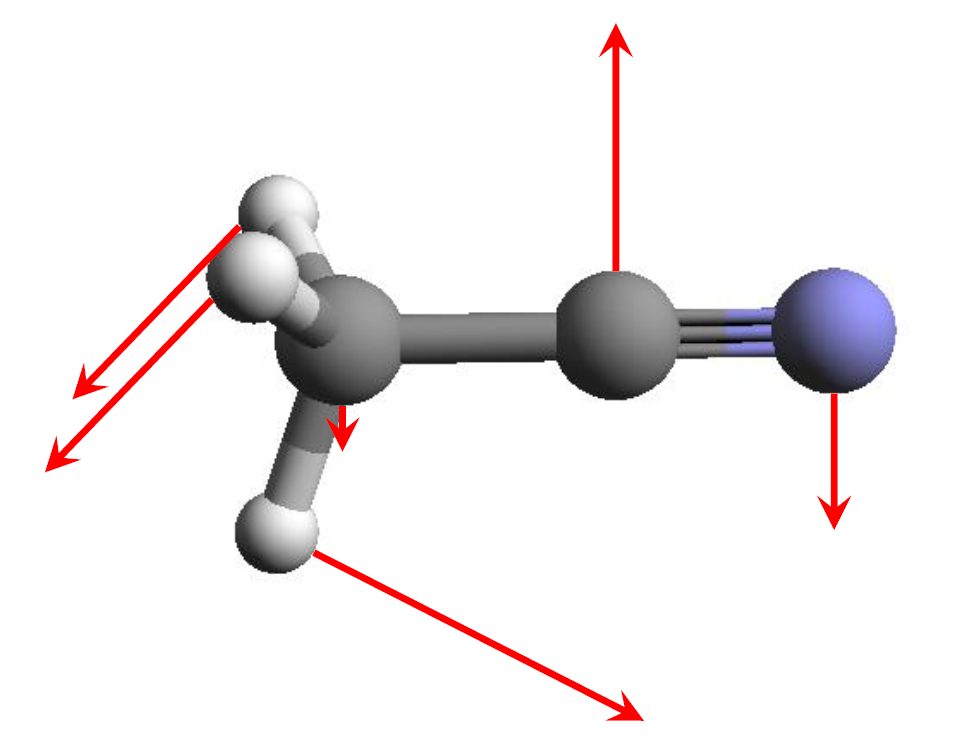}
 \end{center}
  \caption{Model of the methyl cyanide molecule with the $\nu _8$ displacement vectors. 
           The C atoms are shown in gray, the H atoms in light gray, and the N atom in blue. 
           The $a$-axis is along the CCN atoms and is also the symmetry axis. 
           The lengths of the $\nu _8$ displacement vectors are exaggerated.}
  \label{CH3CN_molecule}
 \end{figure}

%%%%%%%%%%%%%%%%%%%%%%%%%%%%%%%%%%%%%%%%%%%%%%%%%%%%%%%%%%%%%%%%%%%%%
%%%%%%%%%%%%%%%%%%%%%%%%%%%%%%%%%%%%%%%%%%%%%%%%%%%%%%%%%%%%%%%%%%%%%

Methyl cyanide is a $C_{\rm 3v}$ symmetric top molecule of the prolate type 
with $A \gg B$ as the light H atoms are the only ones not on the symmetry axis 
of CH$_3$CN, see \autoref{CH3CN_molecule}. 
The general selection rules are $\Delta k - \Delta l \equiv 0$ mod 3, with 
$k$ being the product of $K$ and the sign of $l$ for $l \neq 0$ and with an 
unsigned $k$ indicating an $l = 0$ (sub-) state.

\begin{figure}
  \includegraphics[angle=0,width=9cm]{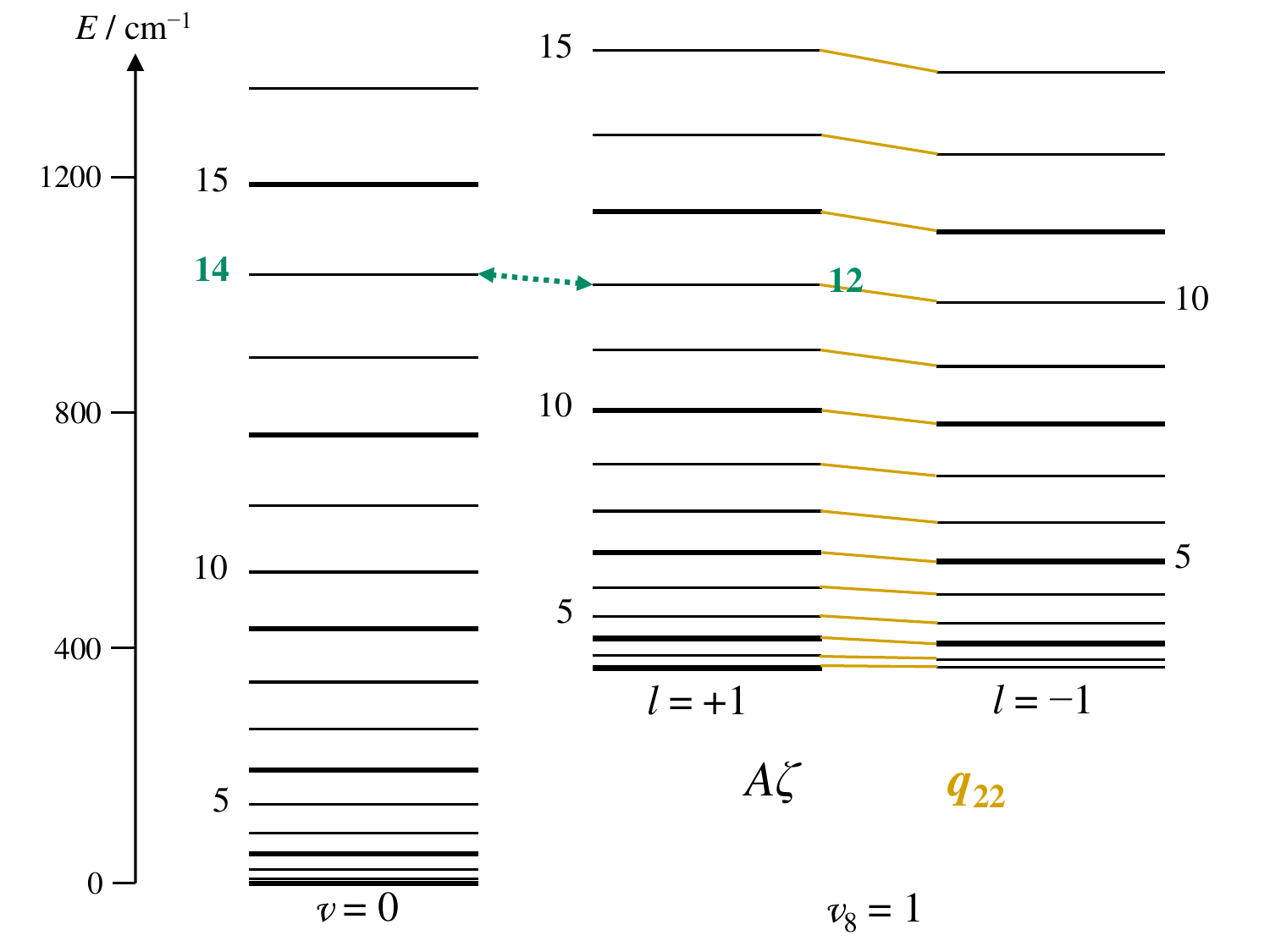}
  \caption{$J = K$ energy levels of methyl cyanide $\varv = 0$ on the left 
   and the doubly degenerate $\varv _8 = 1$ on the right; with the latter separated 
   into their $l = +1$ and $l = -1$ substates. 
   Levels with $k - l \equiv 0$~mod 3 have A symmetry while all others have E symmetry. 
   A symmetry levels are shown with thicker lines. 
   The Coriolis interaction between the two $l = \pm1$ substates of $\varv _8 = 1$ 
   with the lowest order parameter $A\zeta$ shifts $l = +1$ down in energy and 
   $l = -1$ up in energy, causing levels in $\varv _8 = 1$ having the same 
   $K - l$ to be close in energy, facilitating $q_{22}$ interaction. 
   Note the proximity of $K = 12$ of $\varv _8 = 1^{+1}$ to $K = 14$ of $\varv = 0$. 
   The $K = 0$, 1, and 2 levels of $\varv _8 = 1^{+1}$ are so close in energy that their 
   lines in the figure appear as one.}
\label{v8_0_1_egy}
\end{figure}

The CH$_3$CN ground state dipole moment of 3.92197~(13)~D \cite{MeCN-dipole} 
leads to very strong transitions, which obey $\Delta K = 0$ selection rules. 
The Boltzmann peak at 300~K is near 600~GHz at $J$ values slightly above 30, 
and the intensities near 1090~GHz are only 15\% of the peak intensities. 
$\Delta K = 3$ transitions gain intensity through centrifugal distortion effects, 
but are usually too weak to be observed. 
Therefore, the purely axial parameters $A$ (or $A - B$), $D_K$, etc. cannot be 
determined by rotational spectroscopy, unless perturbations are present. 
Rovibrational spectroscopy yields, strictly speaking, the differences $\Delta A$ 
(or $\Delta (A - B)$), $\Delta D_K$, etc. from single state analyses. Thus, the 
ground state axial parameters cannot be determined from such fits either. 
In the case of CH$_3$CN, they were determined through analyses of three IR bands 
involving two doubly degenerate vibrational modes, $\nu_8$, $\nu_7 + \nu_8$, and 
$\nu_7 + \nu_8 - \nu_8$ \cite{MeCN_DeltaK=3_1993}, and were improved through perturbations 
\cite{MeCN_v8le2_2015,MeCN_up2v4eq1_etc_2021}.
The large value of $A$, $\sim$5.27~cm$^{-1}$, leads to a rapid increase in 
the ground state rotational energy with $K$, as shown in \autoref{v8_0_1_egy} 
for the $J = K$ lowest rotational levels.

%%%%%%%%%%%%%%%%%%%%%%%%%%%%%%%%%%%%%%%%%%%%%%%%%%%%%%%%%%%%%%%%%%%%%
%%    Figure 2    %%%%%%%%%%%%%%%%%%%%%%%%%%%%%%%%%%%%%%%%%%%%%%%%%%%
%%%%%%%%%%%%%%%%%%%%%%%%%%%%%%%%%%%%%%%%%%%%%%%%%%%%%%%%%%%%%%%%%%%%%

%%%%%%%%%%%%%%%%%%%%%%%%%%%%%%%%%%%%%%%%%%%%%%%%%%%%%%%%%%%%%%%%%%%%%
%%%%%%%%%%%%%%%%%%%%%%%%%%%%%%%%%%%%%%%%%%%%%%%%%%%%%%%%%%%%%%%%%%%%%

Substitution of an atom on the $C_{\rm 3v}$ symmetry axis does not affect the 
equilibrium $A$ rotational parameter in the Born-Oppenheimer approximation. 
Therefore, it was assumed that the $A_0$ ground state rotational parameters 
do not change upon such substitution either. While this assumption 
probably does not hold strictly, deviations are most likely small.

The lowest CH$_3$CN vibration is the doubly degenerate $\nu _8$ at 365.024~cm$^{-1}$ 
\cite{MeCN_v8le2_2015,MeCN_up2v4eq1_etc_2021}. It is commonly described as the 
CCN bending mode, but the displacement vectors in \autoref{CH3CN_molecule} suggest that 
it is better described as a combination of the CCN bending mode with additional CH$_3$ wagging character. 
The displacement vectors are from a B3LYP \cite{Becke_1993,LYP_1988} quantum chemical calculation 
with the aug-cc-pVTZ basis set \cite{cc-pVXZ_1989} employing Gaussian 03 \cite{Gaussian03B}. 
The displacement vectors were lengthend to make the one of the methyl-C better visible.
Strong Coriolis interaction between the $l$ components is frequently observed in 
low-lying degenerate bending modes. The Coriolis parameter $\zeta$ in $\varv _8 = 1$ of 
CH$_3$CN is 0.8775, close to the limiting case of 1. The $K$ levels with $l = +1$ 
are pushed down in energy, and those with $l = -1$ are pushed up with the result 
that levels with the same $K - l$ are close in energy (\autoref{v8_0_1_egy}). 
These levels have the same symmetry and can thus repel each other through $q_{22}$ 
($\Delta k = \Delta l = \pm2$) interaction.

\begin{figure}
  \includegraphics[angle=0,width=9cm]{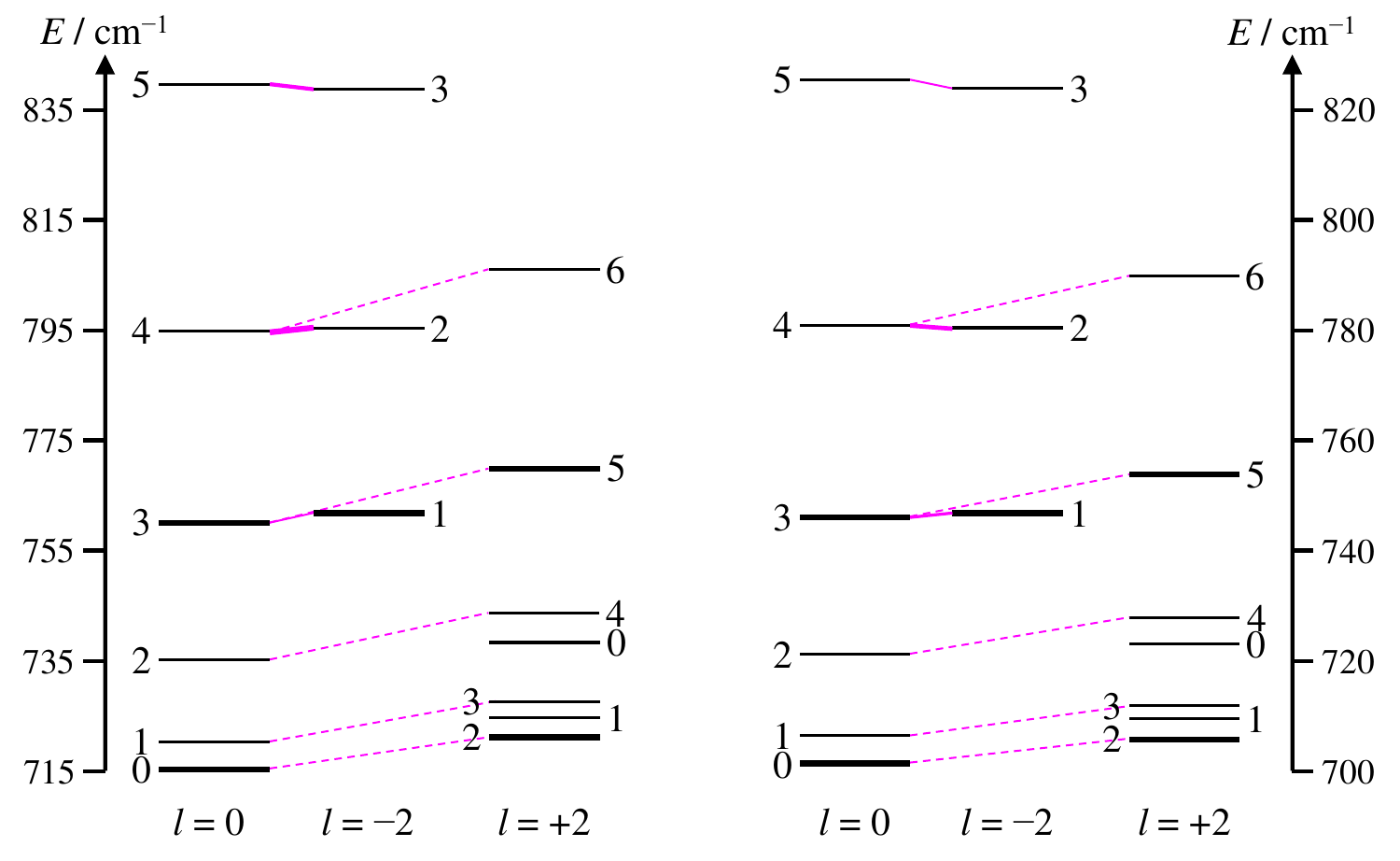}
  \caption{Low-$K$ energy level structure of the triply degenerate 
   $\varv _8 = 2$ of $^{13}$CH$_3$CN on the left and of CH$_3$$^{13}$CN on the right, 
   separated into their $l = 0$, $l = -2$, and $l = +2$ substates. 
   Here, the hypothetical $J = 0$ level energies are shown. 
   The Coriolis interaction between the two $l = \pm2$ substates of $\varv _8 = 2$ 
   shifts the $l = +2$ levels down in energy and the $l = -2$ levels up in energy, 
   causing levels having the same $K - l$ to be close in energy, as indicated by magenta lines, 
   facilitating again $q_{22}$ interaction, similer to $\varv _8 = 1$.
   Note that $K = 4$ of $l = 0$ and $K = 2$ of $l = -2$ are very close 
   in energy with the former being slightly lower than the latter in the case of $^{13}$CH$_3$CN 
   whereas it is opposite in the case of CH$_3$$^{13}$CN. 
   Note also that for $l = +2$ $K = 1$, 2, and 3 are lower in energy than $K = 0$.}
\label{v8_eq_2_low-K_egy}
\end{figure}

There are three $l$ components in the case of $\varv _8 = 2$ with $l = 0$ and $l = \pm2$ 
with origins at 716.750 and 739.148~cm$^{-1}$, respectively, for CH$_3$CN (\autoref{v8_eq_2_low-K_egy}). 
The effective strength of the Coriolis interaction between the $l = +2$ and $l = -2$ levels 
is two times that between the $l = \pm1$ components in $\varv _8 = 1$, causing levels with 
the same $K - l$ for all three $l$ components to be close in energies to a different degree, 
as indicated by the magenta lines in \autoref{v8_eq_2_low-K_egy} for low-$K$ values.

%%%%%%%%%%%%%%%%%%%%%%%%%%%%%%%%%%%%%%%%%%%%%%%%%%%%%%%%%%%%%%%%%%%%%
%%    Figure 3    %%%%%%%%%%%%%%%%%%%%%%%%%%%%%%%%%%%%%%%%%%%%%%%%%%%
%%%%%%%%%%%%%%%%%%%%%%%%%%%%%%%%%%%%%%%%%%%%%%%%%%%%%%%%%%%%%%%%%%%%%

%%%%%%%%%%%%%%%%%%%%%%%%%%%%%%%%%%%%%%%%%%%%%%%%%%%%%%%%%%%%%%%%%%%%%
%%%%%%%%%%%%%%%%%%%%%%%%%%%%%%%%%%%%%%%%%%%%%%%%%%%%%%%%%%%%%%%%%%%%%

The three H atoms are equivalent in CH$_3$CN as well as in isotopologs with substitutions 
on the $C_{\rm 3v}$ axis. This leads to levels with $k - l \equiv 0$ mod 3 having A symmetry 
and all other levels having E symmetry. A level transitions, except those with $K = 0$ of 
(sub-) states with $l \equiv 0$ mod 3, are twice as strong as E levels with about the same energy. 
However, these A levels may in theory be split further into A$_1$ and A$_2$ (or A$_+$ and A$_-$) levels. 
Such splitting was not resolved for CH$_3$CN and its $^{13}$C isotopologs in $k = 3$ of $\varv = 0$ 
and even much less so for higher $k$. However, $k = +1$ of $\varv _8 = 1$ is widely split by the 
$q_{22}$ interaction the two lines are often designated as $k = l = -1$ and $k = l = +1$, respectively; 
the splitting of $k = -2$ of $\varv _8 = 1$ is usually not resolved, 
and levels with higher $K$ even less so. In the case of $\varv _8 = 2$, the splitting in 
$k = +2$ by the $q_{22}$ interaction is frequently resolvable, as can be seen in 
\autoref{v8_eq_2_low-K} close to the upper frequency edge of each trace. 
The splitting in other $k$ levels is by far too small to be resolved.

%%%%%%%%%%%%%%%%%%%%%%%%%%%%%%%%%%%%%%%%%%%%%%%%%%%%%%%%%%%%%%%%%%%%%
%%%%%%%%%%%%%%%%%%%%%%%%%%%%%%%%%%%%%%%%%%%%%%%%%%%%%%%%%%%%%%%%%%%%%
\section{CALCULATION AND FITTING OF THE SPECTRA}
\label{spcat-spfit}
%%%%%%%%%%%%%%%%%%%%%%%%%%%%%%%%%%%%%%%%%%%%%%%%%%%%%%%%%%%%%%%%%%%%%
%%%%%%%%%%%%%%%%%%%%%%%%%%%%%%%%%%%%%%%%%%%%%%%%%%%%%%%%%%%%%%%%%%%%%

The calculation of the $^{13}$CH$_3$CN and CH$_3$$^{13}$CN rotational spectra and 
the fitting of the data were carried out with Pickett's SPCAT and SPFIT programs \cite{spfit_1991}. 
The programs were written as general purpose programs to be capable of treating 
asymmetric top rotors with spins and with vibration-rotation interaction. 
They have evolved considerably with time because of added features, 
in particular special considerations for symmetric top or linear molecules or 
for higher symmetry cases \cite{spins-in-spfit_2004,editorial_Herb-Ed,intro_JPL-catalog}.

We evaluated rotational, centrifugal, and hyperfine structure (HFS) parameters of the ground state 
as common for all vibrational states. Some of the data were measured or reported with partial 
or fully resolved HFS, but the majority of the data were not affected by HFS. 
Therefore, all states were defined twice, with and without HFS. 
The HFS parameter $eqQ\eta$, also designated as $eQq_2$, may require some explanation. 
It is the nuclear quadrupole analog of the $l$-type doubling parameter $q$ \cite{HCN_010_MBER_1970,OCS-isos_010_MBER_1974}. 
It corresponds to an off-diagonal $\chi_{bb} - \chi_{cc}$ in an asymmetric top molecule 
and may be better known as $eQq_2$ from the rotational spectroscopy of $\pi$ radicals. 
Vibrational changes $\Delta X = X_{\rm i} - X_0$ to the ground vibrational state were fit 
for excited vibrational states, where $X$ represents a parameter and $X_{\rm i}$ and $X_0$ 
represent the parameter in excited and ground vibrational states, respectively. 
This is very similar to several early studies on CH$_3$CN \cite{MeCN_nu4_nu7_3nu8_1993,pentade_1994}, 
identical to our previous reports \cite{MeCN_v8le2_2015,MeCN_isos_v8_rot_2016,MeCN_up2v4eq1_etc_2021}, 
and rather convenient because vibrational corrections $\Delta X$ are usually small with 
respect to $X$, especially in the case of low order parameters. 
Moreover, this procedure offers the opportunity to constrain vibrational corrections, 
for example the distortion parameters of $\varv_8 = 2$ to twice those of $\varv_8 = 1$, 
thus reducing the amount of independent spectroscopic parameters further.
New parameters in the fit were chosen carefully by searching for the parameter that 
reduces the rms error of the fit the most. We tried to assess if the value of a new 
parameter is reasonable in the context of related parameters and tried to omit or 
constrain parameters whose values changed considerably in a fit or had relatively 
large uncertainties. Care was also taken that a new parameter is reasonable with 
respect to quantum numbers of newly added transition frequencies or that it can 
account for systematic residuals.

The spectroscopic parameters used in the present analyses are standard symmetric rotor 
parameters defined and designated in a systematic way. The designation of the interaction 
parameters in particular may differ considerably with respect to other publications, and 
there may be small changes in the details of their definitions. Therefore, we give a 
summary of the interaction parameters in the following. Fermi and other anharmonic 
interaction parameters are designated with a plain $F$ and are used in the same way 
irrespective of a $\Delta l = 0$ or $\Delta l = 3$ interaction because the SPFIT and SPCAT 
programs use only $l = 0$ and $\pm1$. The parameters $G_b$ and $F_{ac}$ are first and 
second order Coriolis-type parameters, respectively, of $b$-symmetry, i.e., they are 
coefficients of $iJ_b$ and $(J_aJ_c + J_cJ_a)/2$, respectively. The parameters $G_a$ and 
$F_{bc}$ are defined equivalently. The interacting states are given in parentheses 
separarated by a comma; the degree of excitation of a fundamental and the $l$ quantum number 
are given as superscripts separated by a comma if necessary. Rotational corrections to these 
three types of parameters are designated with $J$ and $K$ subscripts, respectively, as is 
usually the case. There may also be $\Delta k = \Delta l = \pm2$ corrections (i.e. $J_+^2 + J_-^2$; 
where $J_{\pm} = J_a \pm iJ_b$) to these parameters; they are indicated by a subscript 2. 
Higher order corrections with $\Delta k = \Delta l = \pm4$ etc. are defined and indicated equivalently. 
Additional aspects relevant to the spectroscopy of CH$_3$CN were detailed 
earlier \cite{MeCN_v8le2_2015,MeCN_isos_v8_rot_2016,MeCN_up2v4eq1_etc_2021}. 
Further, and more general information on SPFIT and SPCAT is available in \cite{spfit_Novick_2016} and in \cite{spfit_Drouin_2017} 
and in the Fitting Spectra section\cite{Fitting-Spectra} of the Cologne Database for Molecular Spectroscopy, CDMS \cite{CDMS_2001,CDMS_2005,CDMS_2016}. 
It is worthwhile mentioning that there is only one $K = 0$ in states having $l = \pm1$ or $l = \pm2$, 
and the assignment to the respective $l$ component is arbitrary in theory. 
In SPCAT and SPFIT, $K = 0$ is associated with $l = -1$ and with $l = +2$.

Initial spectroscopic parameters for $^{13}$CH$_3$CN and CH$_3$$^{13}$CN $\varv _8 \le 2$ 
were evaluated by taking the respective parameters from our $\varv _8 = 1$ study of 
isotopic methyl cyanide \cite{MeCN_isos_v8_rot_2016}. Estimated parameters were added or 
adjusted based on our latest results on CH$_3$CN \cite{MeCN_up2v4eq1_etc_2021}. 
Parameters of $^{13}$CH$_3$CN and CH$_3$$^{13}$CN determined in the $\varv = 0$ 
investigation \cite{MeCN_rot_2009} and in the $\varv _8 = 1$ study \cite{MeCN_isos_v8_rot_2016} 
were then fit to the respective $\varv _8 \le 1$ data.

%%%%%%%%%%%%%%%%%%%%%%%%%%%%%%%%%%%%%%%%%%%%%%%%%%%%%%%%%%%%%%%%%%%%%
%%%%%%%%%%%%%%%%%%%%%%%%%%%%%%%%%%%%%%%%%%%%%%%%%%%%%%%%%%%%%%%%%%%%%
\section{LABORATORY SPECTROSCOPIC RESULTS}
\label{lab-results}
%%%%%%%%%%%%%%%%%%%%%%%%%%%%%%%%%%%%%%%%%%%%%%%%%%%%%%%%%%%%%%%%%%%%%
%%%%%%%%%%%%%%%%%%%%%%%%%%%%%%%%%%%%%%%%%%%%%%%%%%%%%%%%%%%%%%%%%%%%%

In the following, we describe our results obtained in the course of the present investigation. 
Each part detailing observations starts with a brief description of the previous data available and those used in this study. 
\autoref{v8eq2-results} deals with the results for $\varv _8 = 2$ of $^{13}$CH$_3$CN and CH$_3$$^{13}$CN, 
the corresponding results for $\varv _8 = 1$ are given in \autoref{v8eq1-results}, 
and the ones for the ground vibrational states in \autoref{veq0-results}. 
Details on the combined fits of these three vibrational states of both isotopomers are presented in \autoref{spec-fitting}, 
and \autoref{doubly-subst-isos-results} contains the results obtained for three doubly substituted CH$_3$CN isotopologs.

%%%%%%%%%%%%%%%%%%%%%%%%%%%%%%%%%%%%%%%%%%%%%%%%%%%%%%%%%%%%%%%%%%%%%
%%    Figure 4    %%%%%%%%%%%%%%%%%%%%%%%%%%%%%%%%%%%%%%%%%%%%%%%%%%%
%%%%%%%%%%%%%%%%%%%%%%%%%%%%%%%%%%%%%%%%%%%%%%%%%%%%%%%%%%%%%%%%%%%%%

\begin{figure}
  \includegraphics[angle=0,width=9cm]{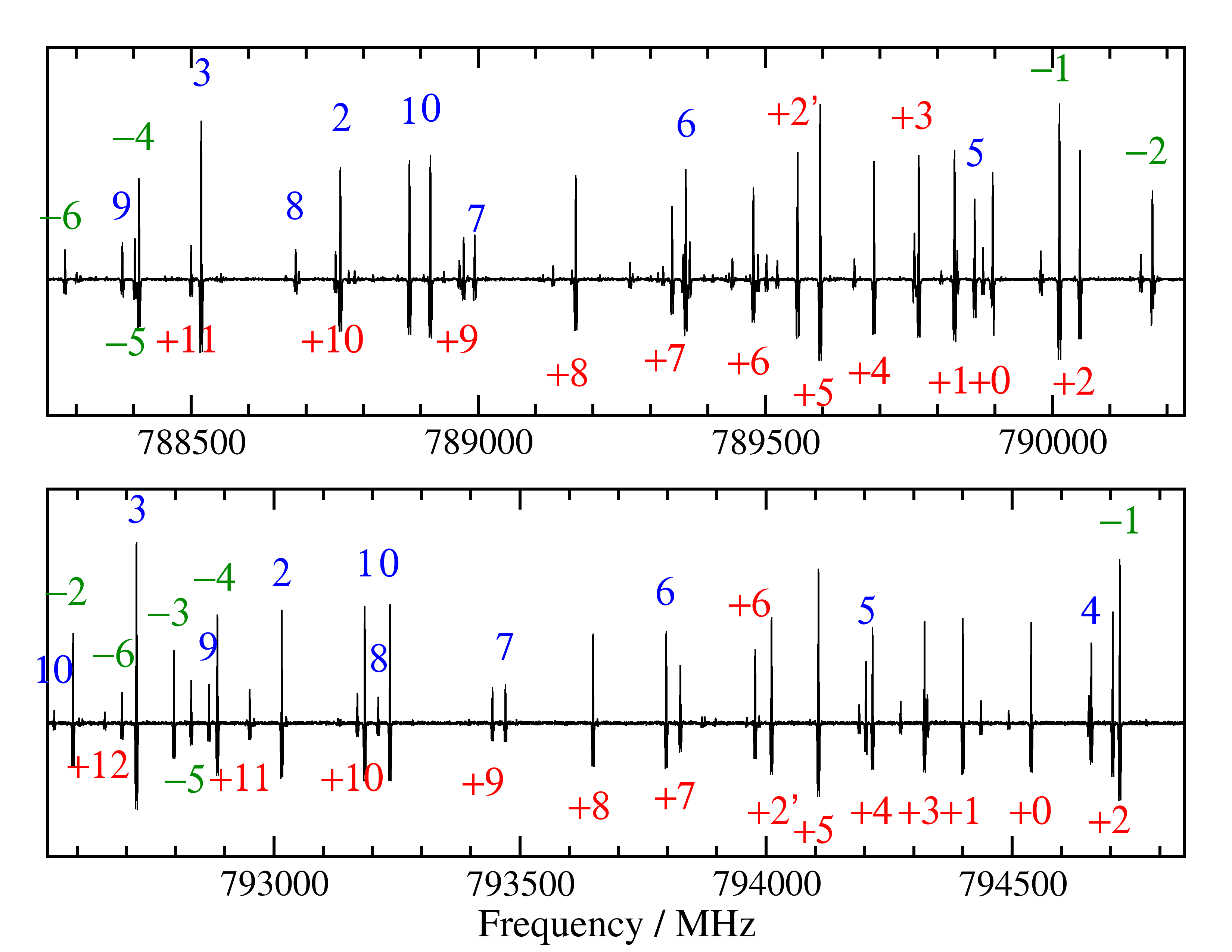}
  \caption{Section of the submillimeter spectrum of $^{13}$CH$_3$CN in the region 
   of the $\varv _8 = 2$ $J = 44 - 43$ transitions on the upper trace and 
   of CH$_3^{13}$CN in the region of $\varv _8 = 2$ $J = 43 - 42$ on the lower trace. 
   The $k$ values of the different $l$ substates are indicated centered above or below the line, 
   for $l = 0$ without a sign, and in different colors; see \autoref{intro-spec} for the definition of $k$. 
   The patterns of both isotopomers are quite similar, in particular with the decrease 
   in frequency for increasing $k$ in the case of $l = +2$ (red); 
   two lines appear for $k = +2$ because the respective levels are split by 
   $q_{22}$ interaction, see also \autoref{intro-spec}. 
   Note, however, that $k = -2$ (in green) occurs at the high frequency edge on the upper trace, 
   whereas it occurs near the low frequency part on the lower trace. 
   Similarly, $k = 4$ (in blue) is near the high frequency edge on the lower trace, whereas 
   it is slightly off the low frequency edge of the upper trace.
   This is caused by the near-degeneracy of $k = -2$ and $k = 4$ together 
   with the $q_{22}$ interaction, see also \autoref{v8_eq_2_low-K_egy}.}
\label{v8_eq_2_low-K}
\end{figure}

%%%%%%%%%%%%%%%%%%%%%%%%%%%%%%%%%%%%%%%%%%%%%%%%%%%%%%%%%%%%%%%%%%%%%
%%    Figure 5    %%%%%%%%%%%%%%%%%%%%%%%%%%%%%%%%%%%%%%%%%%%%%%%%%%%
%%%%%%%%%%%%%%%%%%%%%%%%%%%%%%%%%%%%%%%%%%%%%%%%%%%%%%%%%%%%%%%%%%%%%

\begin{figure}
  \includegraphics[angle=0,width=9cm]{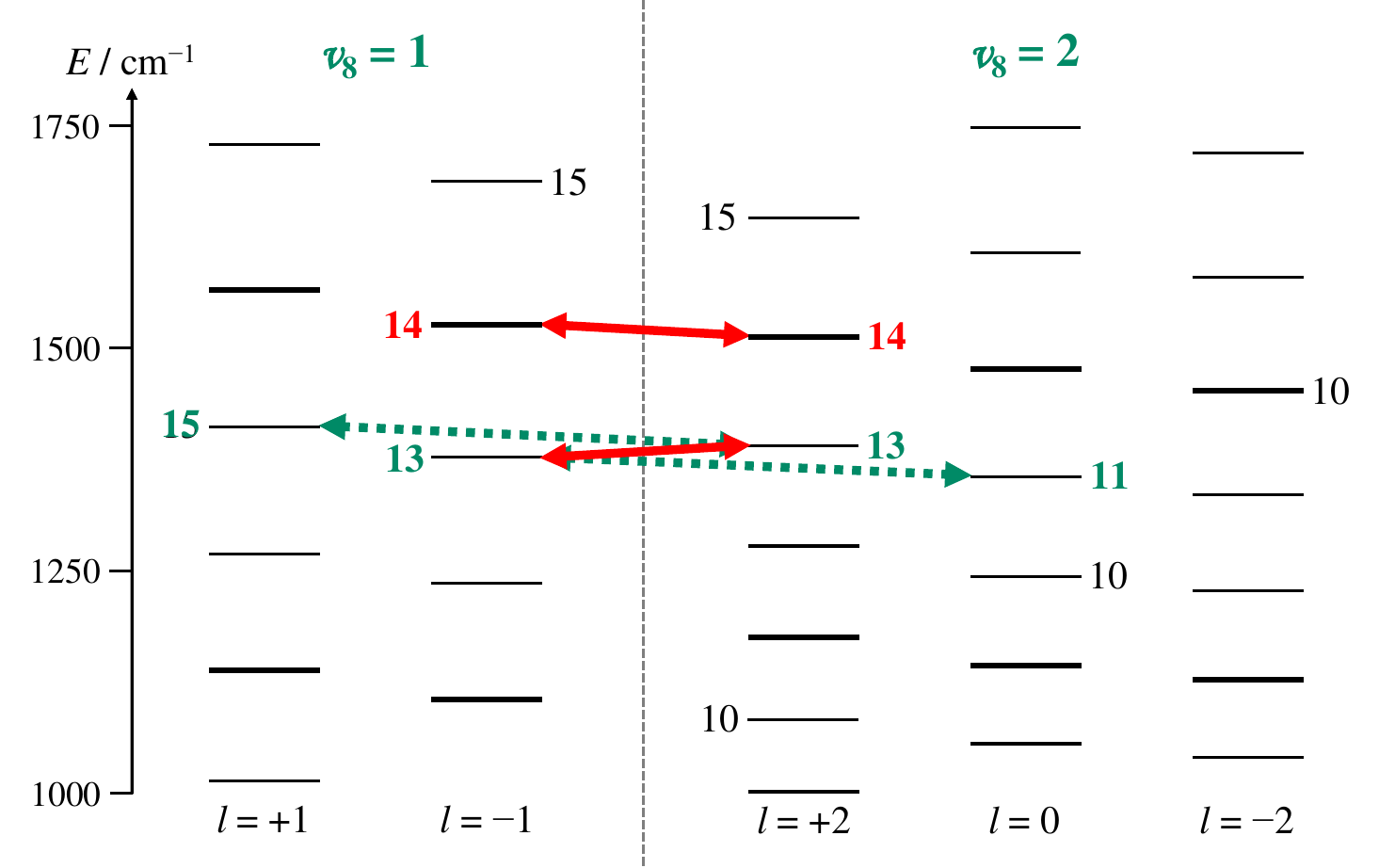}
  \caption{Section of the $J = K$ energy levels of $^{13}$CH$_3$CN $\varv _8 = 1$ on the left 
   and $\varv _8 = 2$ on the right displaying the $K$ levels around their interactions. 
   Levels having the same $K - l$ in $\varv _8 = 1$ are close in energy, 
   and this occurs also within $\varv _8 = 2$. Energy levels with $K = 13$ or 14 
   in $\varv _8 = 1^{-1}$ and in $\varv _8 = 2^{+2}$ are close in energy, 
   giving rise to Fermi-type interaction ($\Delta K = \Delta J = 0$, $\Delta l = \pm3$). 
   In addition, $K = 13$ of $\varv _8 = 1^{-1}$ and $K = 11$ of $\varv _8 = 2^{0}$ 
   get close in energy, as do $K = 15$ of $\varv _8 = 1^{+1}$ and $K = 13$ of $\varv _8 = 2^{+2}$; 
   see also \autoref{v8eq2-results}.}
\label{v8_1_2_egy}
\end{figure}

%%%%%%%%%%%%%%%%%%%%%%%%%%%%%%%%%%%%%%%%%%%%%%%%%%%%%%%%%%%%%%%%%%%%%
%%%%%%%%%%%%%%%%%%%%%%%%%%%%%%%%%%%%%%%%%%%%%%%%%%%%%%%%%%%%%%%%%%%%%
\subsection{\texorpdfstring{$\varv _8 = 2$}{v8 = 2} States of \texorpdfstring{$^{13}$CH$_3$CN and CH$_3$$^{13}$CN}{(13)CH3CN and CH3(13)CN}}
\label{v8eq2-results}
%%%%%%%%%%%%%%%%%%%%%%%%%%%%%%%%%%%%%%%%%%%%%%%%%%%%%%%%%%%%%%%%%%%%%
%%%%%%%%%%%%%%%%%%%%%%%%%%%%%%%%%%%%%%%%%%%%%%%%%%%%%%%%%%%%%%%%%%%%%
The only published $\varv _8 = 2$ transition frequencies of $^{13}$CH$_3$CN and CH$_3$$^{13}$CN 
were reported, to the best of our knowledge, in ref.~\citenum{13C-MeCN_rot_1988}. 
These comprise $J = 1 - 0$ to $3 - 2$ up to 55.5~GHz. 
After fitting these data, $J = 6 - 5$ transition frequencies from earlier recordings 
in natural isotopic composition \cite{MeCN_v8le2_2015,MeCN_isos_v8_rot_2016} 
around 107.8~GHz and around 110.9~GHz, respectively, could be assigned. 
These data were omitted later because of new and more accurate measurements.

After fitting of these data, assignments above 788 and 774~GHz were made 
in spectral recordings of samples enriched in $^{13}$CH$_3$CN and CH$_3$$^{13}$CN, respectively. 
Large fractions of two low $J$ transitions of these recordings, 
$J = 44 - 43$ for $^{13}$CH$_3$CN and $43 - 42$ for CH$_3$$^{13}$CN, 
are shown in \autoref{v8_eq_2_low-K} with the $k$ values indicated above or below each line. 
The patterns of both isotopomers exhibit similarities that are most obvious 
for the $l = +2$ components in red for which the frequencies decrease with increasing $k$. 
The $k = +2$ lines only seem to be displaced because of their splitting caused by the $q_{22}$ interaction, 
the average positions are between $k = +1$ and $+3$. 
The $l = 0$ patterns in blue are more irregular, but some resemblance to each other is noticeable. 
However, there are pronounced differences also. 
Whereas $k = 4$ of CH$_3$$^{13}$CN in the lower trace occurs near the high frequency edge, 
it is below the low frequency edge of the upper trace in the case of $^{13}$CH$_3$CN. 
The $k = -2$ lines in green mirror this appearance, as it is at the upper frequency edge in the upper trace 
for $^{13}$CH$_3$CN, whereas it is near the lower frequency edge in the lower trace for CH$_3$$^{13}$CN. 
As one can see in \autoref{v8_eq_2_low-K_egy}, $k = 4$ and $k = -2$ are very close in energy. 
However, $k = -2$ is $\sim$0.36~cm$^{-1}$ higher than $k = 4$ for low-$J$ values of $^{13}$CH$_3$CN, 
whereas it is $\sim$0.41~cm$^{-1}$ lower in the case of CH$_3$$^{13}$CN. 
The $q_{22}$ interaction and the proximity of these levels cause them to repel each other, 
shifting the upper level up in energy and the lower level down. 
This particular interaction, and also the less pronounced ones for other, in particular low $k$ values of $l = 0$ and $l = -2$, 
cause irregularities in the $k$ line patterns in the spectrum and permit accurate evaluations of the $l = 0$ and $l = \pm2$ energy differences.

After having established the energy differences between the $l$ components of $\varv _8 = 2$ 
for both $^{13}$C containing isotopomers, assignments of transitions in broader frequency regions 
were straightforward for most $k$ up to fairly high values. These broad scans covered many $J$ 
up to near the upper frequency limit near 1090~GHz and reached $J = 60 - 59$ and $59 - 58$ 
for $^{13}$CH$_3$CN and CH$_3$$^{13}$CN, respectively.

%%%%%%%%%%%%%%%%%%%%%%%%%%%%%%%%%%%%%%%%%%%%%%%%%%%%%%%%%%%%%%%%%%%%%
%%    Figure 6    %%%%%%%%%%%%%%%%%%%%%%%%%%%%%%%%%%%%%%%%%%%%%%%%%%%
%%%%%%%%%%%%%%%%%%%%%%%%%%%%%%%%%%%%%%%%%%%%%%%%%%%%%%%%%%%%%%%%%%%%%

\begin{figure}
  \includegraphics[angle=0,width=9cm]{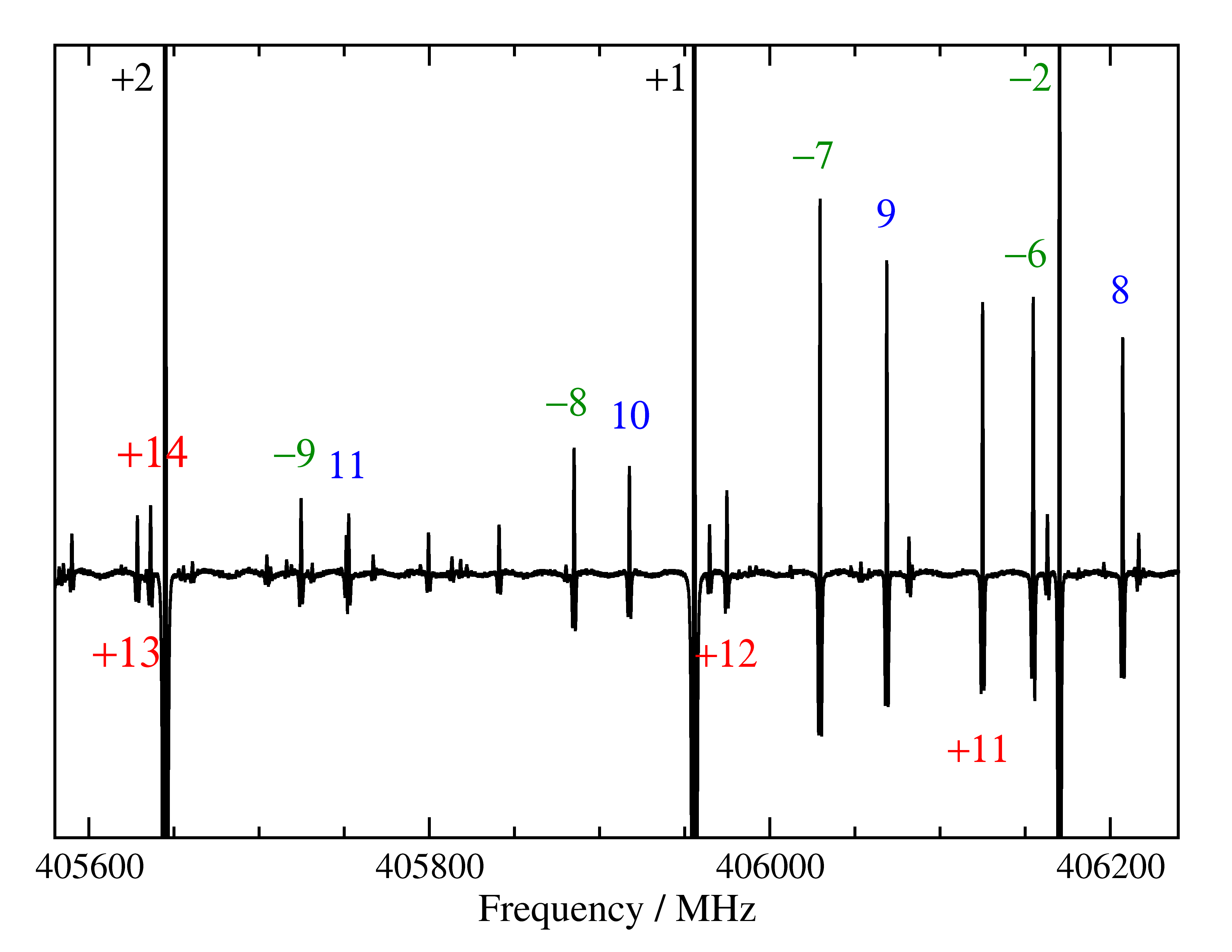}
  \caption{Section of the submillimeter spectrum of CH$_3$$^{13}$CN in the region 
   of $\varv _8 = 2$ $J = 22 - 21$ displaying the effect of the $\Delta l = \pm3$ 
   Fermi-type resonance. Transitions with $k = +11$, 9, and $-7$ occur at 
   successively lower frequencies, as do $k = +12$, 10, and $-8$ lower still. 
   However, $k = +13$ is shifted down in frequency by more than 150~MHz, 
   whereas $k = +14$ is shifted up to a frequency slightly higher than $k = +13$. 
   Two clipped lines labeled with $k = +2$ and $+1$ belong to $\varv _8 = 1$ of CH$_3$$^{13}$CN.}
\label{v8_eq_2_high-K}
\end{figure}

%%%%%%%%%%%%%%%%%%%%%%%%%%%%%%%%%%%%%%%%%%%%%%%%%%%%%%%%%%%%%%%%%%%%%
%%%%%%%%%%%%%%%%%%%%%%%%%%%%%%%%%%%%%%%%%%%%%%%%%%%%%%%%%%%%%%%%%%%%%

As expected, deviations from the calculations were somewhat larger at higher $K$ in the 
$l = +2$ substate, mainly because of a Fermi-type resonance with $\varv _8 = 1^{-1}$ at $K = 13$ and 14, 
as can be seen in \autoref{v8_1_2_egy}. 
The energy levels of $^{13}$CH$_3$CN and CH$_3$CN are very similar mainly because 
of the very small isotopic $\nu _8$ shift of $\sim$0.25~cm$^{-1}$, whereas the shift is nearly 8~cm$^{-1}$ for CH$_3$$^{13}$CN.
This resonance had to be considered already in our 
investigation of $\varv _8 = 1$ of $^{13}$CH$_3$CN and CH$_3$$^{13}$CN in natural isotopic composition 
\cite{MeCN_isos_v8_rot_2016} even though $k$ values only reached $-11$ and $-10$, respectively. 
The energies of $K = 13$ in low $J$ of $\varv _8 = 1^{-1}$ are $\sim$15~cm$^{-1}$ lower than $K = 13$ 
in $\varv _8 = 2^{+2}$ in the case of $^{13}$CH$_3$CN, and the energy differences increase for higher $J$. 
The energies of $K = 14$ in low $J$ of $\varv _8 = 1^{-1}$ are $\sim$13~cm$^{-1}$ lower than $K = 14$ 
in $\varv _8 = 2^{+2}$, but the energy differences decrease to higher $J$, but are still $\sim$7~cm$^{-1}$ at $J = 70$. 
The situation is qualitatively similar for CH$_3$$^{13}$CN, but the energy difference is only 
$\sim$5~cm$^{-1}$ at low $J$ for $K = 13$, increasing to $\sim$11~cm$^{-1}$ at $J = 70$, 
whereas it is nearly $\sim$24~cm$^{-1}$ at low $J$ for $K = 14$, decreasing to still 
$\sim$17~cm$^{-1}$ at $J = 70$. 
The effect of this resonance can be seen in \autoref{v8_eq_2_high-K}. 
Lines with the same values of $K - l$ occur at similar frequencies at these higher $K$ values 
with the $l = +2$ line being higher than the $ l = 0$ line, which in turn is higher than the $l = -2$ line. 
This holds for $k = +11$, 9, and $-7$ as well as for $k = +12$, 10, and $-8$. 
However, $k = +13$ is shifted down well below $k = -9$, and $k = +14$ is shifted up to slightly above $k = +13$.

As indicated in \autoref{v8_1_2_egy}, there are two additional resonances between $\varv _8 = 2$ and 
$\varv _8 = 1$; one is between $k = 11$ and $k = -13$, with an avoided crossing between $J = 68$ and 69 
for $^{13}$CH$_3$CN, slightly too high in $J$ to be covered in the present investigation, 
and the other is between $k = +13$ and $k = +15$ with an avoided crossing between $J = 59$ and 60 for the same isotopolog. 
Transitions involving these $J$ values as well as many lower ones and up to $J = 61$ are in our data set for one or both $k$. 
The perturbations exceed 100~MHz for the most affected transitions and are still larger than 10~MHz in transitions with $J'' \ge 54$ in our data set. 
Both avoided crossings also occur for the main isotopolog, but at slightly lower $J$, at $J = 60$ instead of 68/69 
in the first case and at $J = 52$/53 instead of 59/60 in the second case \cite{MeCN_v8le2_2015}.  
No avoided crossing occurs in the $J$ range covered for CH$_3$$^{13}$CN as the energy differences still exceed 6~cm$^{-1}$ 
at $J = 70$ for both resonances. The perturbed lines are shifted by less than 1~MHz for this isotopolog.

Low-$K$ lines, in particular those exhibiting the strongest $q_{22}$ perturbations, as well as high-$K$ lines of the $l = +2$ substate 
were recorded individually for the transitions above 770~GHz that were not covered by broad scans. 
At lower frequencies, $J = 2 - 1$ to $6 - 5$ were recorded, which exhibit HFS split transitions almost throughout. 
In addition, transitions of some $J$ were recorded between $\sim$160 and $\sim$630~GHz.

After having completed the analyses of the present spectral recordings, we inspected spectral recordings of methyl cyanide 
in natural isotopic composition covering most of the 1083 to 1200~GHz region that were taken at JPL much earlier. 
We expected at least some lines to have sufficiently good S/N because we had made $^{13}$CH$_3$$^{13}$CN assignments in that region \cite{MeCN_rot_2009}. 
The strongest lines of $\varv _8 = 2$ of both isotopomers exhibited quite good S/N between 1085 and 1148~GHz; 
the lines were too weak at still higher frequencies.

Assigned transitions extend to $k = +15$ for both $^{13}$CH$_3$CN and CH$_3$$^{13}$CN, fitting reasonably well for most of the former, and less so for the latter isotopomer. 
Poorly fitting lines appearing not to be blended where kept in the fit, but with sufficient MHz values added to the uncertainties, essentially weighting out these lines. 
Lines turning out to be perturbed, may be weighted in if the perturbation will have been resolved. 
Very small numbers of assignments of $k = +16$ and $+17$ exist for both isotopomers, but none of them fit well at present.

In the case of the $l = 0$ substate, assigned transitions reach $K = 12$ plus two $K = 13$ lines in the case of $^{13}$CH$_3$CN, 
with lines fitting well for the most part up to $K = 7$ for both isotopomers. 
All but one transitions with $K = 9$ fit well for CH$_3$$^{13}$CN. The deviations for $K \ge 10$ are probably caused for the most part 
by a Fermi-type resonance with $\varv _8 = 3^{+3}$ which were calculated to be strongest at $K = 15$ for the main isotopolog \cite{MeCN_v8le2_2015}. 
Transitions with $K = 8$ may be perturbed by $K = 5$ of $\varv _4 = 1$, for which an avoided crossing occurs at $J = 57$ for CH$_3$CN \cite{MeCN_up2v4eq1_etc_2021}. 
The small and regular deviations in $K = 9$ of $^{13}$CH$_3$CN are not easily explained at present. 
Untangling of these perturbations is beyond the scope of the present investigation, as it would probably not suffice 
to estimate spectroscopic parameters for $\varv _4 = 1$, $\varv _7 = 1$, and $\varv _8 = 3$ from values of the main isotopolog \cite{MeCN_up2v4eq1_etc_2021}. 
Instead, it is probably necessary to include $\varv _7 = 1$ and $\varv _8 = 3$ data of the main isotopic species and redetermine its parameters 
prior to a meaningful estimation of parameters of the singly substituted $^{13}$C isotopomers. 
In addition, we may have to include at least some experimental data for the three higher vibrational states of the $^{13}$C isotopomers 
to adjust $B$ and possibly some additional low order parameters along with some of the multitude of interaction parameters. 
We point out that vibrational energies of $\varv _4 = 1$ and $\varv _7 = 1$ of methyl cyanide isotopologs 
are fairly well known from low-resolution IR measurements \cite{FF_Duncan_1978}.

We assigned up to $K = 10$ in the $l = -2$ substate plus one $K = 11$ line for $^{13}$CH$_3$CN. 
The lines fit well up to $K = 5$; $K = 6$ interacts with $K = 5$ of $\varv _4 = 1$, this may also apply to the next higher $K$. 
Effects of a Fermi-type resonance between $\varv _8 = 2^{-2}$ and $\varv _8 = 3^{+1}$, that is strongest at $K = 12$ and 13, 
may already affect $K = 7$ of $\varv _8 = 2^{-2}$, almost certainly $K = 8$ and higher. 
The line lists of both isotopomers are available as Supporting Information and have also been deposited in the CDMS as detailed in the Data Availability Statement.

%%%%%%%%%%%%%%%%%%%%%%%%%%%%%%%%%%%%%%%%%%%%%%%%%%%%%%%%%%%%%%%%%%%%%
%%%%%%%%%%%%%%%%%%%%%%%%%%%%%%%%%%%%%%%%%%%%%%%%%%%%%%%%%%%%%%%%%%%%%
\subsection{\texorpdfstring{$\varv _8 = 1$}{v8 = 1} States of \texorpdfstring{$^{13}$CH$_3$CN and CH$_3$$^{13}$CN}{(13)CH3CN and CH3(13)CN}}
\label{v8eq1-results}
%%%%%%%%%%%%%%%%%%%%%%%%%%%%%%%%%%%%%%%%%%%%%%%%%%%%%%%%%%%%%%%%%%%%%
%%%%%%%%%%%%%%%%%%%%%%%%%%%%%%%%%%%%%%%%%%%%%%%%%%%%%%%%%%%%%%%%%%%%%
The previous $\varv _8 = 1$ transition frequencies of $^{13}$CH$_3$CN and CH$_3$$^{13}$CN 
were taken from Ref~\citenum{MeCN_isos_v8_rot_2016} and originated entirely from that work. 
Earlier limited low frequency data \cite{13C-MeCN_rot_1988} were already omitted in that work 
because then new $J = 3 - 2$ data were obtained with much higher accuracies. 
The transition frequencies with HFS splitting involve $J = 3 - 2$ and $6 - 5$ for both isotopomers and $5 - 4$ for CH$_3$$^{13}$CN.

The new measurements cover many $k$ of several $J$ values as well as recordings 
of individual lines in particular at higher $k$ values, similar to the $\varv _8 = 2$ recordings. 
The $k$ values reached $+16$, $+17$, $-15$, and $-16$ in several cases and the next two pairs of $k$ only in 0 to 2 cases. 
Emphasis was put on measuring transitions with $k$ involved in $\varv _8 = 2$ interactions.

%%%%%%%%%%%%%%%%%%%%%%%%%%%%%%%%%%%%%%%%%%%%%%%%%%%%%%%%%%%%%%%%%%%%%
%%    Figure 7    %%%%%%%%%%%%%%%%%%%%%%%%%%%%%%%%%%%%%%%%%%%%%%%%%%%
%%%%%%%%%%%%%%%%%%%%%%%%%%%%%%%%%%%%%%%%%%%%%%%%%%%%%%%%%%%%%%%%%%%%%

\begin{figure}
  \includegraphics[angle=0,width=9cm]{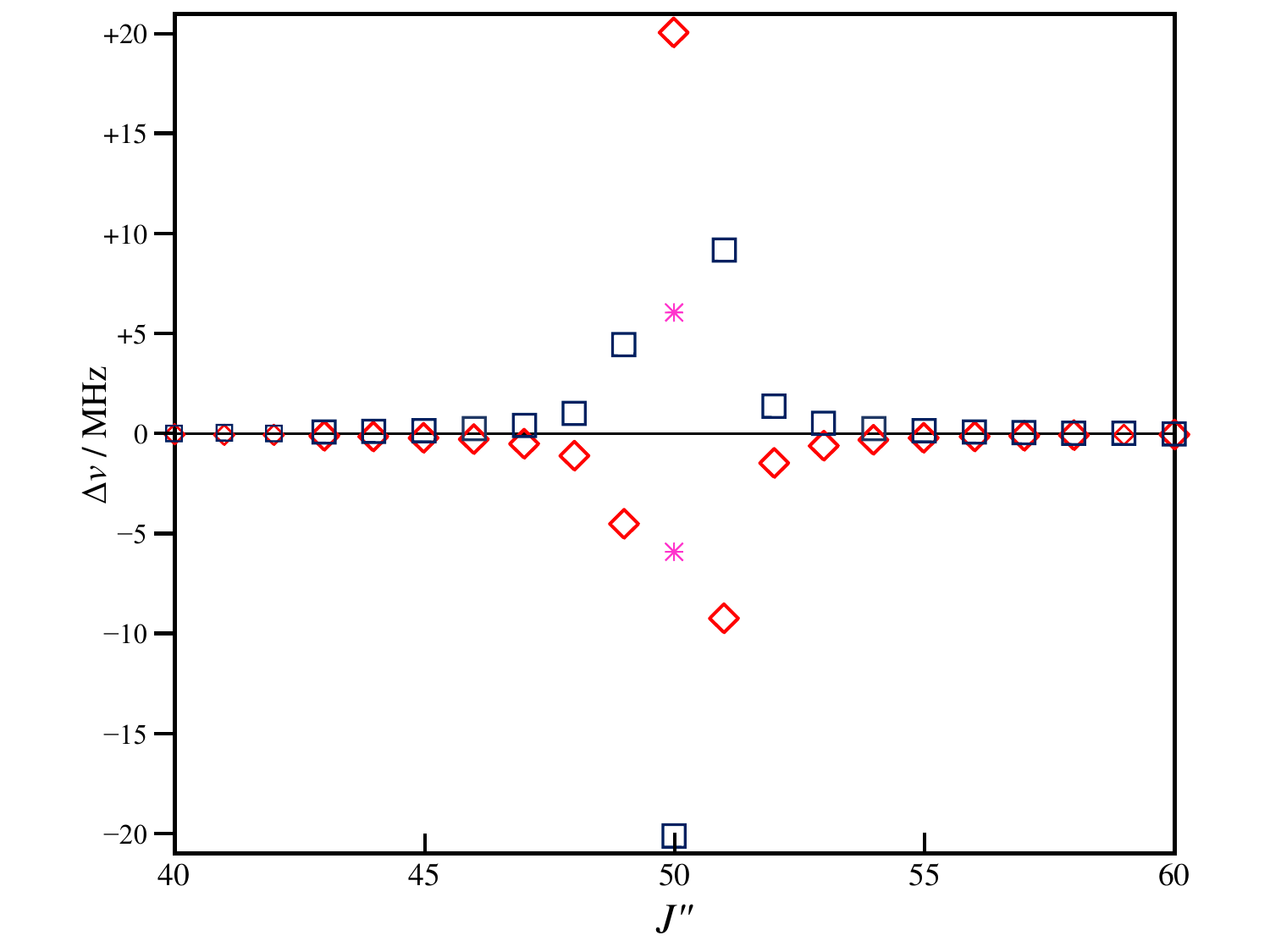}
  \caption{Perturbation plot of $^{13}$CH$_3$CN. 
   Differences $\Delta \nu$ between the $J' - J''$ transition frequencies calculated from the final spectroscopic parameters and 
   those calculated with the interaction parameter $F_2(0,8^{\pm1})$ set to zero are shown for 
   $\varv = 0$ $K = 14$ (blue squares) and $\varv _8 = 1^{+1}$ $K = 12$ (red diamonds). 
   Larger symbols indicate transitions in the final fit. The very weak cross ladder transitions (magenta) were not observed.}
\label{perturbation-plot}
\end{figure}

%%%%%%%%%%%%%%%%%%%%%%%%%%%%%%%%%%%%%%%%%%%%%%%%%%%%%%%%%%%%%%%%%%%%%
%%    Figure 8    %%%%%%%%%%%%%%%%%%%%%%%%%%%%%%%%%%%%%%%%%%%%%%%%%%%
%%%%%%%%%%%%%%%%%%%%%%%%%%%%%%%%%%%%%%%%%%%%%%%%%%%%%%%%%%%%%%%%%%%%%

\begin{figure}
  \includegraphics[angle=0,width=9cm]{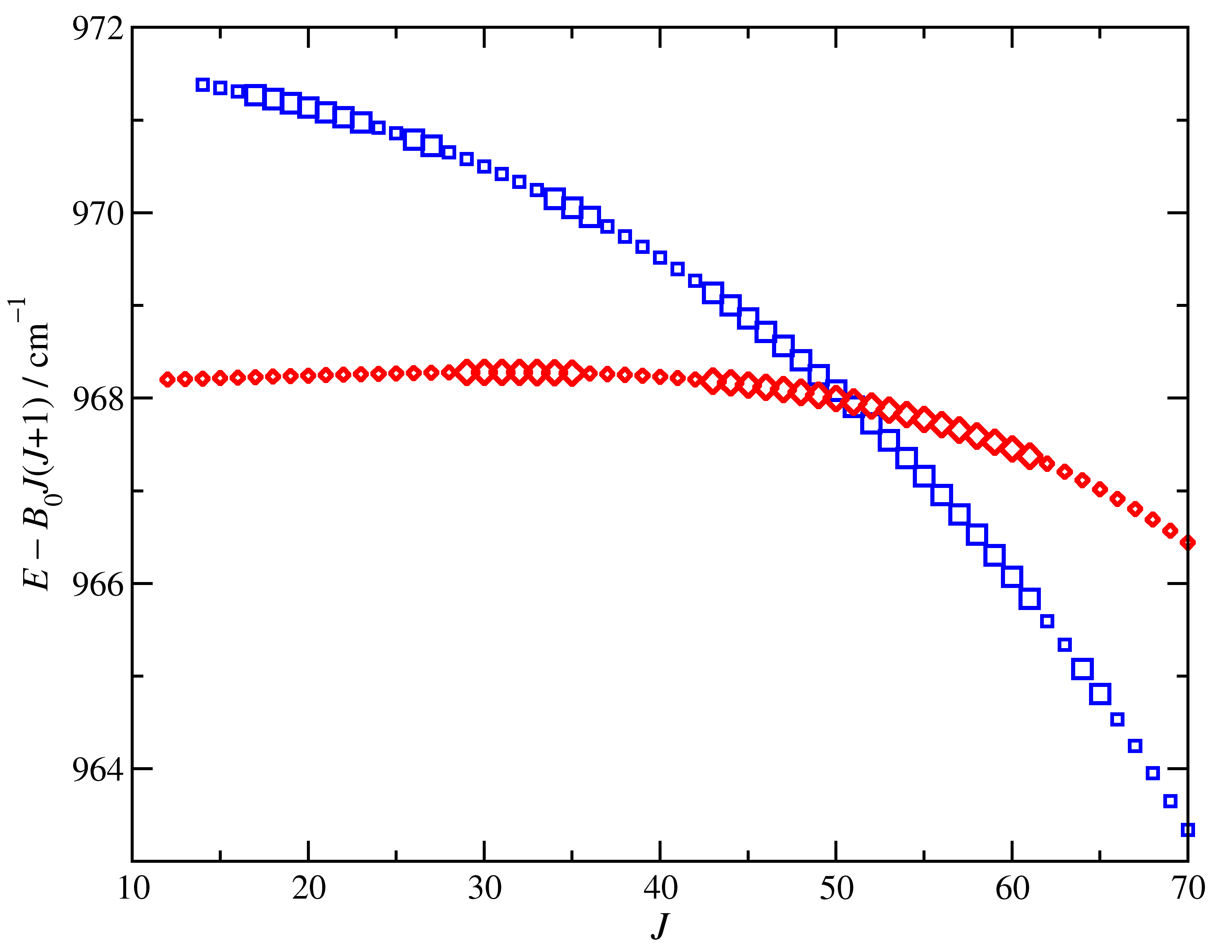}
  \caption{Reduced energy plot of $^{13}$CH$_3$CN in the region of 
   $\varv = 0$ $K = 14$ (blue squares) and $\varv _8 = 1^{+1}$ $K = 12$ (red diamonds) 
   displaying the resonant interaction at $J = 51$. Larger symbols indicate 
   levels that have been accessed by experimental transitions in the final fit. 
   This interaction probes the $\varv _8 = 1$ vibrational energy as 364.77~cm$^{-1}$.}
\label{14_0-12_3-reso}
\end{figure}

%%%%%%%%%%%%%%%%%%%%%%%%%%%%%%%%%%%%%%%%%%%%%%%%%%%%%%%%%%%%%%%%%%%%%
%%%%%%%%%%%%%%%%%%%%%%%%%%%%%%%%%%%%%%%%%%%%%%%%%%%%%%%%%%%%%%%%%%%%%

While the proximity of $k = 14$ in $\varv = 0$ and $k = +12$ in $\varv _8 = 1$ was known, see \autoref{v8_0_1_egy}, 
it appeared initially that the energy differences would be too large for both $^{13}$C isotopomers to access a level crossing. 
However, some regularly increasing deviations appeared in $k = +12$ starting at $J = 44 - 43$ near 785~GHz in the case of $^{13}$CH$_3$CN. 
Inspection of the energy file revealed a level crossing to occur at $J = 51$. 
The deviations in $k = 14$ of $\varv = 0$ mirrored these deviations; the largest ones were 20~MHz for the $J = 51 - 50$ 
transitions, as can be seen in \autoref{perturbation-plot}. 
\autoref{14_0-12_3-reso} displays part of the reduced energy plot involving these two $k$ 
values, demonstrating that they only get very close in energy near $J$ of 51. 
The level crossing is at $J = 42$/43 for CH$_3$CN \cite{MeCN_v8le2_2015}, whereas it is calculated to be at $J = 91$/92 for CH$_3$$^{13}$CN, 
well outside the covered $J$ range and probably much too weak to be observable, as the Boltzmann peak at 300~K is at $J = 32 - 31$.

%%%%%%%%%%%%%%%%%%%%%%%%%%%%%%%%%%%%%%%%%%%%%%%%%%%%%%%%%%%%%%%%%%%%%
%%%%%%%%%%%%%%%%%%%%%%%%%%%%%%%%%%%%%%%%%%%%%%%%%%%%%%%%%%%%%%%%%%%%%
\subsection{\texorpdfstring{$\varv = 0$}{v = 0} States of \texorpdfstring{$^{13}$CH$_3$CN and CH$_3$$^{13}$CN}{(13)CH3CN and CH3(13)CN}}
\label{veq0-results}
%%%%%%%%%%%%%%%%%%%%%%%%%%%%%%%%%%%%%%%%%%%%%%%%%%%%%%%%%%%%%%%%%%%%%
%%%%%%%%%%%%%%%%%%%%%%%%%%%%%%%%%%%%%%%%%%%%%%%%%%%%%%%%%%%%%%%%%%%%%
The previous $\varv = 0$ transition frequencies of $^{13}$CH$_3$CN and CH$_3$$^{13}$CN 
were taken from ref.~\citenum{MeCN_rot_2009} with then new data between 312 and 1193~GHz. 
The line lists contained earlier data from ref.~\citenum{MeCN-isos_rot_1996} with frequencies for the hyperfine split 
$J = 1 - 0$ transitions along with unsplit transition frequencies between 294 and 608~GHz. 
Almost all of these data were retained, except a few transition frequencies of CH$_3$$^{13}$CN near 440~GHz. 
Also included were some millimeter data \cite{MeCN-Lille_1977} which were eventually omitted 
in the present work because of newly recorded, much more accurate data at similar frequencies. 
Finally, accurate transition frequencies of the $J = 2 - 1$ transition of CH$_3$$^{13}$CN 
\cite{MeCN-12-13b_2-1} were also included previously and are retained in the present work.

The new measurements from this study cover transitions with HFS splitting between 35 and 111~GHz 
($J = 2 - 1$ to $6 - 5$) with data of the $J = 4 - 3$ transition of $^{13}$CH$_3$CN near 71.46~GHz 
taken from earlier recordings of a sample in natural isotopic composition because of its better quality. 
Transitions without HFS splitting were recorded between 355 and 1087~GHz, 
reaching $J = 61 - 60$ and $59 - 58$ for $^{13}$CH$_3$CN and CH$_3$$^{13}$CN, respectively, 
covering a fair fraction of $J$ in this region, in particular at higher frequencies. 
The $K$ range was covered as completely as possible, reaching $K = 21$ two times in the case of CH$_3$$^{13}$CN 
and $K = 19$ two times for $^{13}$CH$_3$CN, in both cases for two $J$ values with frequencies slightly above 600~GHz. 
Transitions with $K = 14$ were recorded later for $^{13}$CH$_3$CN between $J = 54 - 53$ to $61 - 60$ 
because of the resonance of $\varv = 0$ and $\varv _8 = 1$ as mentioned in the previous \autoref{v8eq1-results}.

%%%%%%%%%%%%%%%%%%%%%%%%%%%%%%%%%%%%%%%%%%%%%%%%%%%%%%%%%%%%%%%%%%%%%
%%%%%%%%%%%%%%%%%%%%%%%%%%%%%%%%%%%%%%%%%%%%%%%%%%%%%%%%%%%%%%%%%%%%%
\subsection{Fitting of the \texorpdfstring{$^{13}$CH$_3$CN and CH$_3$$^{13}$CN}{(13)CH3CN and CH3(13)CN} Data}
\label{spec-fitting}
%%%%%%%%%%%%%%%%%%%%%%%%%%%%%%%%%%%%%%%%%%%%%%%%%%%%%%%%%%%%%%%%%%%%%
%%%%%%%%%%%%%%%%%%%%%%%%%%%%%%%%%%%%%%%%%%%%%%%%%%%%%%%%%%%%%%%%%%%%%
The $\varv_8 \le 2$ data of each singly $^{13}$C substituted isotopomer were fit together employing Pickett's SPFIT program \cite{spfit_1991}, 
as done in earlier reports on the main isotopic species \cite{MeCN_v8le2_2015,MeCN_up2v4eq1_etc_2021}. 
The equilibrium $A$ value does not change upon substituting one of the atoms on the $C_{\rm 3v}$ axis; 
therefore, we assumed that $A$ in $\varv = 0$ and the $\Delta A$ values of excited vibrational (sub-) states do not change either for such substitutions. 
This is the same procedure as in our $\varv _8 = 1$ study of singly substitued methyl cyanide isotopologs \cite{MeCN_isos_v8_rot_2016} 
for $\varv = 0$, but slightly different for the excited states, as we assumed earlier $\Delta (A - B)$ to be the same 
for a given vibrational (sub-) state upon subsitution on the $C_{\rm 3v}$ axis. 
Either assumption is probably reasonable, but unlikely to hold strictly; deviations of order of 1 or a few MHz are deemed to be possible. 
The isotopic ratios of higher order spectroscopic parameters were estimated to scale with appropriate powers of $B$, 
as in an ealier study \cite{MeCN_isos_v8_rot_2016}; for example, $D$ scales with the ratio of $B^2$.  
In order to improve the initial higher order parameter estimates, isotopic ratios of the low-order parameters $A\zeta$, $q$, and $F$ 
were also applied to the respective higher order parameters; thus $q_J$ scales with the ratios of $q$ and $B$. 
Initially, we assumed rotational corrections of $A\zeta$ and $q$ to be the same in $\varv _8 = 1$ and 2; this holds rather often to first order. 
We point out that there are vibrational factors of these parameters which are different between $\varv _8 = 1$ and 2 in SPFIT. 
Vibrational changes $\Delta X$ of distortion parameters in $\varv _8 = 2$ were initially constrained to be two times those in $\varv _8 = 1$, 
which is frequently fulfilled to first order if $l$-dependent changes (of degenerate vibrational states) can be neglected.

%%%%%%%%%%%%%%%%%%%%%%%%%%%%%%%%%%%%%%%%%%%%%%%%%%%%%%%%%%%%%%%%%%%%%
%%    Table 1    %%%%%%%%%%%%%%%%%%%%%%%%%%%%%%%%%%%%%%%%%%%%%%%%%%%%
%%%%%%%%%%%%%%%%%%%%%%%%%%%%%%%%%%%%%%%%%%%%%%%%%%%%%%%%%%%%%%%%%%%%%

\begin{table*}
\begin{center}
\caption{Spectroscopic Parameters or Vibrational Changes ($\Delta$) with Respect to the Ground Vibrational States\textsuperscript{\emph{a}} of $^{13}$CH$_3$CN and CH$_3$$^{13}$CN 
         in $\varv_8 = 2$ from a Combined Fit of $\varv_8 \le 2$ Data and Comparison with Previous Data\textsuperscript{\emph{b}}.}
\label{parameter_v8_eq_2}
% \smallskip
% \renewcommand{\arraystretch}{1.10}
\resizebox{0.83\linewidth}{!}{
\begin{tabular}[t]{lr@{}lr@{}llr@{}lr@{}l}
\hline 
          & \multicolumn{4}{c}{$^{13}$CH$_3$CN} & & \multicolumn{4}{c}{CH$_3$$^{13}$CN}  \\
\cline{2-5} \cline{7-10}
Parameter & \multicolumn{2}{c}{This work} & \multicolumn{2}{c}{Previous\textsuperscript{\emph{b}}} & & \multicolumn{2}{c}{This work} & \multicolumn{2}{c}{Previous\textsuperscript{\emph{b}}} \\
\hline
$\Delta E(8^{2^2} - 8^{2^0})$              &     22&.928605~(33)                         &     22&.37      & &     21&.792148~(16)                          &     21&.91     \\
$\Delta (A - B)$                           & $-$185&.58                                  & $-$187&.4       & & $-$187&.74                                   & $-$187&.4      \\
$\Delta B$                                 &     52&.100413~(32)                         &     52&.4115    & &     52&.219257~(44)                          &     51&.856    \\
$\Delta D_K \times 10^3$                   &  $-$20&.4\textsuperscript{\emph{c}}         &       &         & &  $-$20&.4\textsuperscript{\emph{c}}          &       &        \\
$\Delta D_{JK} \times 10^3$                &      1&.36866~(292)                         &       &         & &      1&.45386~(225)                          &       &        \\
$\Delta D_J \times 10^6$                   &    211&.432~(7)                             &       &         & &    202&.613~(25)                             &       &        \\
$\Delta H_{KJ} \times 10^6$                &      0&.970~(62)                            &       &         & &      2&.487~(25)                             &       &        \\
$\Delta H_{JK} \times 10^9$                &     13&.10~(46)                             &       &         & &     12&.76~(4)                               &       &        \\
$E(8^{2^2})$                               &    738&.65472~(76)                          &    738&.19      & &    723&.1949~(65)                            &    723&.01     \\
$\Delta (A - B)$                           & $-$260&.705~(120)                           & $-$260&.0       & & $-$255&.946~(79)                             & $-$260&.0      \\
$\Delta B$                                 &     52&.849506~(25)                         &     52&.84335   & &     52&.263148~(32)                          &     52&.2833   \\
$\Delta D_K \times 10^3$                   &  $-$20&.4\textsuperscript{\emph{c}}         &       &         & &  $-$20&.4\textsuperscript{\emph{c}}          &       &        \\
$\Delta D_{JK} \times 10^3$                &      1&.70985~(96)                          &       &         & &      1&.52119~(107)                          &       &        \\
$\Delta D_J \times 10^6$                   &    178&.371~(15)                            &       &         & &    182&.257~(22)                             &       &        \\
$\Delta H_{KJ} \times 10^6$                &   $-$0&.1013~(43)                           &       &         & &   $-$0&.1991~(55)                            &       &        \\
$\Delta H_{JK} \times 10^9$                &   $-$2&.252~(113)                           &       &         & &   $-$4&.829~(159)                            &       &        \\
$\Delta H_J \times 10^{12}$                &    521&.4~(20)                              &       &         & &    536&.7~(26)                               &       &        \\
$\Delta eQq$                               &   $-$0&.0798~(30)\textsuperscript{\emph{c}} &       &         & &   $-$0&.0888~(33)\textsuperscript{\emph{c}}  &       &        \\
$A\zeta$                                   & 138847&.16~(17)                             & 138858&.        & & 139827&.80~(8)                               & 139829&.5      \\
$\eta_K$                                   &     10&.0455~(62)\textsuperscript{\emph{d}} &       &         & &     10&.3882~(285)\textsuperscript{\emph{d}} &       &        \\
$\eta_J$                                   &      0&.375339~(3)                          &       &         & &      0&.392547~(3)                           &       &        \\
$\eta_{KK} \times 10^6$                    & $-$678&.\textsuperscript{\emph{d}}          &       &         & & $-$683&.\textsuperscript{\emph{d}}           &       &        \\
$\eta_{JK} \times 10^6$                    &  $-$31&.939~(22)\textsuperscript{\emph{d}}  &       &         & &  $-$33&.259~(27)\textsuperscript{\emph{d}}   &       &        \\
$\eta_{JJ} \times 10^6$                    &   $-$2&.2659~(23)\textsuperscript{\emph{d}} &       &         & &   $-$2&.2689~(42)\textsuperscript{\emph{d}}  &       &        \\
$\eta_{JKK} \times 10^9$                   &      2&.20\textsuperscript{\emph{d}}        &       &         & &      2&.22\textsuperscript{\emph{d}}         &       &        \\
$\eta_{JJK} \times 10^9$                   &      0&.5475~(56)\textsuperscript{\emph{d}} &       &         & &      0&.6462~(70)\textsuperscript{\emph{d}}  &       &        \\
$q$                                        &     16&.70353~(18)                          &     16&.7382    & &     18&.17787~(9)                            &     18&.1412   \\
$q_{K} \times 10^3$                        &   $-$2&.089~(12)\textsuperscript{\emph{d}}  &       &         & &    $-$2&.698\textsuperscript{\emph{d}}       &       &        \\
$q_{J} \times 10^6$                        &  $-$64&.550~(35)                            &       &         & &   $-$68&.183~(34)                            &       &        \\
$q_{JK} \times 10^9$                       &     90&.06~(94)\textsuperscript{\emph{d}}   &       &         & &      79&.11~(94)\textsuperscript{\emph{d}}   &       &        \\
$q_{JJ} \times 10^{12}$                    &    161&.2~(43)                              &       &         & &     259&.4~(46)                              &       &        \\
$F(8^{\pm1},8^{2,\mp2})$                   &  52777&.6~(27)                              &  51647&.~(1303) & &  50584&.5~(51)                               &  54567&.~(744) \\
$F_K(8^{\pm1},8^{2,\mp2})$                 &   $-$6&.                                    &   $-$6&.        & &   $-$6&.                                     &   $-$6&.       \\
$F_J(8^{\pm1},8^{2,\mp2})$                 &   $-$0&.35815~(43)                          &   $-$0&.359     & &   $-$0&.35460~(41)                           &   $-$0&.370    \\
$F_{JJ}(8^{\pm1},8^{2,\mp2}) \times 10^6$  &      1&.700~(74)                            &      1&.58      & &      1&.783~(115)                            &      1&.70     \\
$F_2(8^{\pm1},8^{2,0}) \times 10^3$        &  $-$52&.58~(5)\textsuperscript{\emph{e}}    &       &         & &  $-$55&.37~(102)\textsuperscript{\emph{e}}   &       &        \\
$F_2(8^{\pm1},8^{2,\pm2}) \times 10^3$     & $-$105&.16~(10)\textsuperscript{\emph{e}}   &       &         & & $-$110&.74~(204)\textsuperscript{\emph{e}}   &       &        \\
$F_{2,J}(8^{\pm1},8^{2,0}) \times 10^6$    &   $-$0&.88\textsuperscript{\emph{e}}        &       &         & &   $-$0&.98\textsuperscript{\emph{e}}         &       &        \\
$F_{2,J}(8^{\pm1},8^{2,\pm2}) \times 10^6$ &   $-$1&.76\textsuperscript{\emph{e}}        &       &         & &   $-$1&.96\textsuperscript{\emph{e}}         &       &        \\
rms error                                  &      0&.877                                 &      0&.766     & &      0&.891                                  &      0&.808    \\
\hline
\end{tabular}
}
\end{center}
\textsuperscript{\emph{a}} All parameters in MHz units except $E(8^{2^2} - 8^{2^0})$ and $E(8^{2^2})$ in cm$^{-1}$; the rms errors of the fits are unitless. 
Vibrational changes $\Delta X$ are defined as $X_{\rm i} - X_0$. Numbers in parentheses are one standard deviation in units of the least significant figures. 
Parameters without quoted uncertainties were estimated and kept fixed in the analyses. 
\textsuperscript{\emph{b}} Reference~\citenum{MeCN_isos_v8_rot_2016}. 
\textsuperscript{\emph{c}} The corresponding $\varv_8 = 2$  and $\varv_8 = 1$ parameters of each isotopomer were constrained to a 2~:~1 ratio; 
see also \autoref{spec-fitting}. 
\textsuperscript{\emph{d}} The corresponding $\varv_8 = 2$  and $\varv_8 = 1$ parameters of each isotopomer were constrained to a 1~:~1 ratio; 
see also \autoref{spec-fitting}. 
\textsuperscript{\emph{e}} A 1~:~2 ratio was constrained for $F_2(8^{\pm1},8^{2,0})$ and $F_2(8^{\pm1},8^{2,\pm2})$ and 
also for their distortion corrections $F_{2,J}$, as done earlier for CH$_3$CN\cite{MeCN_up2v4eq1_etc_2021,MeCN_v8le2_2015}.
\end{table*}
%%%%%%%%%%%%%%%%%%%%%%%%%%%%%%%%%%%%%%%%%%%%%%%%%%%%%%%%%%%%%%%%%%%%%
%%%%%%%%%%%%%%%%%%%%%%%%%%%%%%%%%%%%%%%%%%%%%%%%%%%%%%%%%%%%%%%%%%%%%

We searched in each fitting round for a parameter whose floating in the fit improved the rms error of the fit the most. 
The only parameter tried out here that was not used in the fits of CH$_3$CN was $f_{44}$, which connects levels 
differing in $\Delta k = \Delta l = \pm4$. It was retained in some $^{13}$CH$_3$CN fits, but was omitted later because of its relatively large uncertainty. 
In the case of $\varv _8 = 1$ and 2 parameters, usually a parameter was floated with keeping the $\varv _8 = 1$ and 2 constraint 
before the effect of lifting this constraint was tested. 
Sometimes either procedure led to reinstating a constraint for a different parameter because of changes caused by correlation deemed to be too large. 
It appeared to be difficult, however, to avoid entirely much larger vibrational changes than suggested by the constraints. 
This applied in particular to $\Delta H_{KJ}$, $\Delta H_{JK}$, and $\Delta H_J$. 
Attempts to impose constraints led to substantial increases in the rms errors that could not be compensated by other parameters.

Interestingly, it was possible to float one of the two $\Delta (A - B)$ values of $\varv _8 = 2$. 
Noting that $\Delta (A - B)$ of $\varv _8 = 2^2$ of $^{13}$CH$_3$CN is slightly more than 2~MHz larger in magnitude 
than the one derived under the assumption of $\Delta A$ to be the same as the value of CH$_3$CN, we tried to impose this value. 
The rms error of the fit increased by $\sim$10\% and some parameters took less favorable values. 
Keeping this constraint and floating $\Delta (A - B)$ of $\varv _8 = 2^0$ instead, reduced the rms error to about the previous value 
and caused this $\Delta (A - B)$ value to decrease in magnitude by $\sim$2~MHz with the remaining parameters 
fairly similar to those with $\Delta (A - B)$ of $\varv _8 = 2^2$ floated. 
Attempts to float both $\Delta (A - B)$ of $\varv _8 = 2$ decreased both in magnitude by $\sim$8.5~MHz 
with uncertainties of about 5.5~MHz in the case of $^{13}$CH$_3$CN, whereas both increased by $\sim$25~MHz 
with uncertainties of about 10~MHz in the case of CH$_3$$^{13}$CN. 
These findings suggest that the difference in $A - B$ between the two $l$ substates of $\varv _8 = 2$ is well constrained, but the value of each is not quite yet. 
Therefore, we reverted to fitting only $\Delta (A - B)$ of $\varv _8 = 2^2$ and kept the one of $l = 0$ fixed at the estimated value.

The largest shifts by far caused by the HFS parameter $eQq\eta$ occur in the $\Delta F = 0$ HFS components 
of the $k = l = +1$ and $k = l = -1$ transitions of $\varv _8 = 1$ 
and also in their $F = 1 - 0$ components of the respective $J = 2 - 1$ transitions. 
The shifts are essentially equal in magnitude and opposite in sign among the corresponding $k = l = +1$ and $k = l = -1$ HFS components.

The resulting spectroscopic parameters are listed in \autoref{parameter_v8_eq_2}, \autoref{parameter_v8_eq_1}, 
and \autoref{ground-state-parameter} for $\varv _8 = 2$, 1, and for $\varv = 0$, respectively.
The line, parameter, and fit files of both isotopomers are available as Supporting Information 
and have also been deposited in the CDMS as detailed in the Data Availability Statement.

%%%%%%%%%%%%%%%%%%%%%%%%%%%%%%%%%%%%%%%%%%%%%%%%%%%%%%%%%%%%%%%%%%%%%
%%    Table 2    %%%%%%%%%%%%%%%%%%%%%%%%%%%%%%%%%%%%%%%%%%%%%%%%%%%%
%%%%%%%%%%%%%%%%%%%%%%%%%%%%%%%%%%%%%%%%%%%%%%%%%%%%%%%%%%%%%%%%%%%%%

\begin{table*}
\begin{center}
\caption{Spectroscopic Parameters or Vibrational Changes ($\Delta$) Thereof$^a$ of $^{13}$CH$_3$CN and CH$_3$$^{13}$CN 
         in $\varv_8 = 1$ from a Combined Fit of $\varv_8 \le 2$ Data and Comparison with Previous Data$^b$.}
\label{parameter_v8_eq_1}
% \smallskip
% \renewcommand{\arraystretch}{1.10}
\resizebox{0.97\linewidth}{!}{
\begin{tabular}[t]{lr@{}lr@{}llr@{}lr@{}l}
\hline 
          & \multicolumn{4}{c}{$^{13}$CH$_3$CN} & & \multicolumn{4}{c}{CH$_3$$^{13}$CN}  \\
\cline{2-5} \cline{7-10}
Parameter & \multicolumn{2}{c}{This work} & \multicolumn{2}{c}{Previous\textsuperscript{\emph{b}}} & & \multicolumn{2}{c}{This work} & \multicolumn{2}{c}{Previous\textsuperscript{\emph{b}}} \\
\hline
$E(8^1)$                       &    364&.76782~(35)                          &    364&.56           & &    357&.19                                   &    357&.19           \\
$\Delta (A - B)$               & $-$115&.03                                  & $-$115&.930          & & $-$114&.75                                   & $-$115&.930          \\
$\Delta B$                     &     26&.687766~(32)                         &     26&.688102~(131) & &     26&.405025~(35)                          &     26&.404860~(112) \\
$\Delta D_K \times 10^3$       &  $-$10&.2\textsuperscript{\emph{c}}         &  $-$11&.46           & &  $-$10&.2\textsuperscript{\emph{c}}          &  $-$11&.46           \\
$\Delta D_{JK} \times 10^3$    &      0&.95016~(39)                          &      0&.9453~(26)    & &      0&.86956~(45)                           &      0&.8811~(24)    \\
$\Delta D_J \times 10^6$       &     90&.447~(12)                            &     90&.693~(41)     & &     92&.400~(14)                             &     92&.585~(52)     \\
$\Delta H_K \times 10^6$       &     20&.8                                   &     15&.             & &     20&.8                                    &     15&.             \\
$\Delta H_{KJ} \times 10^6$    &      0&.0331~(16)                           &      0&.033          & &      0&.0202~(17)                            &      0&.034          \\
$\Delta H_{JK} \times 10^9$    &      1&.844~(63)                            &      2&.44           & &      1&.647~(81)                             &      2&.54           \\
$\Delta H_J \times 10^{12}$    &    252&.7~(17)                              &    278&.6~(56)       & &    270&.7~(18)                               &    307&.3~(74)       \\
$\Delta L_J \times 10^{15}$    &       &                                     &   $-$2&.35           & &       &                                      &   $-$2&.54           \\
$\Delta eQq$                   &   $-$0&.0399~(15)\textsuperscript{\emph{c}} &   $-$0&.0387         & &   $-$0&.0444~(17)\textsuperscript{\emph{c}}  &   $-$0&.0387         \\
$eQq\eta$                      &      0&.1647~(39)                           &      0&.1519         & &      0&.1562~(54)                            &      0&.1519         \\
$A\zeta$                       & 138859&.445~(37)                            & 138857&.97~(53)      & & 139838&.800~(37)                             & 139839&.99~(39)      \\
$\eta_K$                       &     10&.0455~(62)\textsuperscript{\emph{d}} &     10&.347          & &     10&.3882~(285)\textsuperscript{\emph{d}} &     10&.420          \\
$\eta_J$                       &      0&.371752~(4)                          &      0&.371764~(23)  & &      0&.389217~(4)                           &      0&.389261~(24)  \\
$\eta_{KK} \times 10^6$        & $-$678&.\textsuperscript{\emph{d}}          & $-$835&.             & & $-$683&.\textsuperscript{\emph{d}}           & $-$840&.             \\
$\eta_{JK} \times 10^6$        &  $-$31&.939~(22)\textsuperscript{\emph{d}}  &  $-$32&.87~(40)      & &  $-$33&.259~(27)\textsuperscript{\emph{d}}   &  $-$32&.42~(44)      \\
$\eta_{JJ} \times 10^6$        &   $-$2&.2659~(23)\textsuperscript{\emph{d}} &   $-$2&.1471~(49)    & &   $-$2&.2689~(42)\textsuperscript{\emph{d}}  &   $-$2&.2750~(52)    \\
$\eta_{JKK} \times 10^9$       &      2&.20\textsuperscript{\emph{d}}        &      2&.5            & &      2&.22\textsuperscript{\emph{d}}         &      2&.6            \\
$\eta_{JJK} \times 10^9$       &      0&.5475~(56)\textsuperscript{\emph{d}} &      0&.425~(70)     & &      0&.6462~(70)\textsuperscript{\emph{d}}  &      0&.542~(69)     \\
$q$                            &     16&.80179~(10)                          &     16&.80442~(35)   & &     18&.21027~(10)                           &     18&.21258~(33)   \\
$q_{K} \times 10^3$            &   $-$2&.089~(12)\textsuperscript{\emph{d}}  &   $-$2&.516          & &   $-$2&.698\textsuperscript{\emph{d}}        &   $-$2&.726          \\
$q_{J} \times 10^6$            &  $-$58&.697~(40)                            &  $-$58&.745~(164)    & &  $-$64&.596~(36)                             &  $-$64&.581~(175)    \\
$q_{JK} \times 10^9$           &     90&.06~(94)\textsuperscript{\emph{d}}   &     85&.4            & &     79&.11~(94)\textsuperscript{\emph{d}}    &     95&.2            \\
$q_{JJ} \times 10^{12}$        &    295&.3~(54)                              &    301&.0~(218)      & &    319&.8~(47)                               &    304&.8~(252)      \\
$F_2(0,8^{\pm1}) \times 10^3$  &  $-$56&.298~(15)                            &       &              & &       &                                      &       &              \\
\hline
\end{tabular}
}
\end{center}
\textsuperscript{\emph{a}} All parameters in MHz units except $E(8^1)$ in cm$^{-1}$. 
Vibrational changes $\Delta X$ are defined as $X_{\rm i} - X_0$. Numbers in parentheses are one standard deviation in units of the least significant figures. 
Parameters without quoted uncertainties were estimated and kept fixed in the analyses. 
\textsuperscript{\emph{b}} Reference~\citenum{MeCN_isos_v8_rot_2016}. 
\textsuperscript{\emph{c}} The corresponding $\varv_8 = 2$  and $\varv_8 = 1$ parameters of each isotopomer were constrained to a 2~:~1 ratio; 
see also \autoref{spec-fitting}. 
\textsuperscript{\emph{d}} The corresponding $\varv_8 = 2$  and $\varv_8 = 1$ parameters of each isotopomer were constrained to a 1~:~1 ratio; 
see also \autoref{spec-fitting}. 
\end{table*}

%%%%%%%%%%%%%%%%%%%%%%%%%%%%%%%%%%%%%%%%%%%%%%%%%%%%%%%%%%%%%%%%%%%%%
%%    Table 3    %%%%%%%%%%%%%%%%%%%%%%%%%%%%%%%%%%%%%%%%%%%%%%%%%%%%
%%%%%%%%%%%%%%%%%%%%%%%%%%%%%%%%%%%%%%%%%%%%%%%%%%%%%%%%%%%%%%%%%%%%%

\begin{table*}
\begin{center}
\caption{Ground-State Spectroscopic Parameters\textsuperscript{\emph{a}} (MHz) of $^{13}$CH$_3$CN and CH$_3$$^{13}$CN 
         from a Combined Fit of $\varv_8 \le 2$ Data and Comparison with Previous Data\textsuperscript{\emph{b}}.}
\label{ground-state-parameter}
% \smallskip
% \renewcommand{\arraystretch}{1.10}
\resizebox{0.97\linewidth}{!}{
\begin{tabular}[t]{lr@{}lr@{}llr@{}lr@{}l}
\hline 
          & \multicolumn{4}{c}{$^{13}$CH$_3$CN} & & \multicolumn{4}{c}{CH$_3$$^{13}$CN}  \\
\cline{2-5} \cline{7-10}
Parameter & \multicolumn{2}{c}{This work} & \multicolumn{2}{c}{Previous\textsuperscript{\emph{b}}} & & \multicolumn{2}{c}{This work} & \multicolumn{2}{c}{Previous\textsuperscript{\emph{b}}} \\
\hline
$(A - B)$                       & 149165&.60          & 149165&.69          & & 148904&.56          & 148904&.65          \\
$B$                             &   8933&.309448~(16) &   8933&.309429~(28) & &   9194&.350055~(17) &   9194&.349998~(27) \\
$D_K \times 10^3$               &   2827&.9           &   2831&.            & &   2827&.9           &   2831&.            \\
$D_{JK} \times 10^3$            &    168&.24018~(47)  &    168&.23971~(130) & &    176&.67618~(48)  &    176&.67395~(132) \\
$D_J \times 10^6$               &   3625&.018~(13)    &   3624&.947~(32)    & &   3809&.866~(14)    &   3809&.737~(37)    \\
$H_K \times 10^6$               &    156&.            &    165&.            & &    156&.            &    165&.            \\
$H_{KJ} \times 10^6$            &      5&.7894~(30)   &      5&.8030~(141)  & &      6&.0018~(28)   &      6&.0045~(131)  \\
$H_{JK} \times 10^9$            &    927&.18~(13)     &    927&.19~(86)     & &   1018&.47~(15)     &   1017&.45~(86)     \\
$H_J \times 10^{12}$            & $-$230&.7~(44)      & $-$273&.3~(49)      & & $-$206&.5~(52)      & $-$258&.4~(61)      \\
$L_{KKJ} \times 10^{12}$        & $-$400&.0~(65)      & $-$431&.            & & $-$416&.1~(55)      & $-$444&.            \\
$L_{JK} \times 10^{12}$         &  $-$47&.35~(34)     &  $-$49&.66~(266)    & &  $-$52&.38~(33)     &  $-$49&.75~(236)    \\
$L_{JJK} \times 10^{12}$        &   $-$6&.896~(16)    &   $-$6&.887~(128)   & &   $-$7&.764~(20)    &   $-$7&.214~(141)   \\
$L_J \times 10^{15}$            &   $-$8&.17~(50)     &   $-$2&.76          & &   $-$9&.20~(61)     &   $-$3&.10          \\
$P_{JK} \times 10^{15}$         &      0&.46          &      0&.51          & &      0&.51          &      0&.55          \\
$P_{JJK} \times 10^{18}$        &     48&.            &     49&.            & &     54&.            &     55&.            \\
$eQq$                           &   $-$4&.21804~(152) &   $-$4&.21830~(197) & &   $-$4&.21860~(145) &   $-$4&.21828~(176) \\
$C_{bb} \times 10^3$            &      1&.79          &      1&.792         & &      1&.84          &      1&.844         \\
$(C_{aa} - C_{bb}) \times 10^3$ &   $-$1&.12          &   $-$1&.10          & &   $-$1&.17          &   $-$1&.15          \\
\hline
\end{tabular}
}
\end{center}
\textsuperscript{\emph{a}} Numbers in parentheses are one standard deviation in units of the least significant figures. 
Parameters without quoted uncertainties have been estimated from the main isotopic species and were kept fixed in the fits. 
\textsuperscript{\emph{b}} Reference~\citenum{MeCN_rot_2009}; adjusted in that work to account for slight changes in the parameters of the main isotopolog 
from that work compared to those in ref.~\citenum{MeCN_v8le2_2015}.  
\end{table*}

%%%%%%%%%%%%%%%%%%%%%%%%%%%%%%%%%%%%%%%%%%%%%%%%%%%%%%%%%%%%%%%%%%%%%
%%%%%%%%%%%%%%%%%%%%%%%%%%%%%%%%%%%%%%%%%%%%%%%%%%%%%%%%%%%%%%%%%%%%%
\subsection{Ground Vibrational States of \texorpdfstring{$^{13}$CH$_3$$^{13}$CN, $^{13}$CH$_3$C$^{15}$N, and CH$_3$$^{13}$C$^{15}$N}{(13)CH3(13)CN, (13)CH3C(15)N, and CH3(13)C(15)N}}
\label{doubly-subst-isos-results}
%%%%%%%%%%%%%%%%%%%%%%%%%%%%%%%%%%%%%%%%%%%%%%%%%%%%%%%%%%%%%%%%%%%%%
%%%%%%%%%%%%%%%%%%%%%%%%%%%%%%%%%%%%%%%%%%%%%%%%%%%%%%%%%%%%%%%%%%%%%
The high enrichment of the $^{13}$CH$_3$CN and CH$_3$$^{13}$CN samples means that $^{13}$CH$_3$$^{13}$CN is enriched 
in both samples to about 1.1\% because of the natural $^{13}$C/$^{12}$C ratio at the C atom that was not enriched. 
Moreover, $^{13}$CH$_3$C$^{15}$N and CH$_3$$^{13}$C$^{15}$N are enriched to about 0.38\% 
in the $^{13}$CH$_3$CN and CH$_3$$^{13}$CN samples, respectively, because of the natural $^{15}$N/$^{14}$N ratio. 
Therefore, our recordings covered some $J$ of the respective isotopologs.

%%%%%%%%%%%%%%%%%%%%%%%%%%%%%%%%%%%%%%%%%%%%%%%%%%%%%%%%%%%%%%%%%%%%%
%%    Table 4    %%%%%%%%%%%%%%%%%%%%%%%%%%%%%%%%%%%%%%%%%%%%%%%%%%%%
%%%%%%%%%%%%%%%%%%%%%%%%%%%%%%%%%%%%%%%%%%%%%%%%%%%%%%%%%%%%%%%%%%%%%

\begin{table*}
\begin{center}
\caption{Spectroscopic Parameters\textsuperscript{\emph{a}} (MHz) of Methyl Cyanide Doubly Substituted 
Species and Dimensionless Weighted Standard Deviation wrms.}
\label{parameter_doubly-subst}
% \smallskip
% \renewcommand{\arraystretch}{1.10}
\resizebox{0.97\linewidth}{!}{
\begin{tabular}[t]{lr@{}lr@{}lr@{}lr@{}l}
\hline
 & \multicolumn{4}{c}{$^{13}$CH$_3$$^{13}$CN} & \multicolumn{2}{c}{$^{13}$CH$_3$C$^{15}$N} &  \multicolumn{2}{c}{CH$_3$$^{13}$C$^{15}$N}\\
\cline{2-5}
Parameter       & \multicolumn{2}{c}{This work} & \multicolumn{2}{c}{Previous\textsuperscript{\emph{b}}} &       &             &       &             \\
\hline
$(A - B) \times 10^{-3}$        &    149&.171692      &    149&.171692       &    149&.434530      &    149&.181347      \\
$B$                             &   8927&.281134~(45) &   8927&.281294~(155) &   8659&.892784~(38) &   8919&.231179~(43) \\
$D_K$                           &      2&.8257        &      2&.8257         &      2&.8257        &      2&.8257        \\
$D_{JK} \times 10^3$            &    167&.4232~(12)   &    167&.4204~(177)   &    160&.0445~(12)   &    168&.4062~(11)   \\
$D_J \times 10^3$               &      3&.627251~(32) &      3&.627152~(61)  &      3&.380957~(37) &      3&.556858~(42) \\
$H_K \times 10^6$               &     51&.            &     51&.             &     51&.            &     51&.            \\
$H_{KJ} \times 10^6$            &      5&.7497~(61)   &      5&.50~(42)      &      5&.4037~(77)   &      5&.6355~(76)   \\
$H_{JK} \times 10^6$            &      0&.91905~(25)  &      0&.9159~(35)    &      0&.85813~(36)  &      0&.94615~(29)  \\
$H_J \times 10^{12}$            & $-$292&.0~(65)      & $-$298&.             & $-$227&.3~(76)      & $-$156&.7~(95)      \\
$L_{KKJ} \times 10^{12}$        & $-$427&.            & $-$427&.             & $-$412&.            & $-$419&.            \\
$L_{JK} \times 10^{12}$         &  $-$47&.3           &  $-$47&.3            &  $-$45&.3           &  $-$45&.3           \\
$L_{JJK} \times 10^{12}$        &   $-$6&.7           &   $-$6&.7            &   $-$6&.122         &   $-$6&.916         \\
$L_J \times 10^{15}$            &   $-$1&.800         &   $-$1&.800          &   $-$1&.007         &   $-$1&.007         \\
$P_{JK} \times 10^{18}$         &    450&.            &    450&.             &    477&.             &    526&.           \\
$P_{JJK} \times 10^{18}$        &     48&.0           &     48&.0            &     40&.1            &     50&.0          \\
$eQq$                           &   $-$4&.2218~(53)   &   $-$4&.21319        &       &              &       &            \\
$C_{bb} \times 10^3$            &      1&.73          &      1&.73           &       &              &       &            \\
$(C_{aa} - C_{bb}) \times 10^3$ &   $-$1&.24          &   $-$1&.24           &       &              &       &            \\
wrms                            &      0&.810         &      0&.635          &      0&.848          &      0&.805        \\
\hline
\end{tabular}
}
\end{center}
\textsuperscript{\emph{a}} Numbers in parentheses are one standard deviation in units of the least significant figures. 
Parameters without quoted uncertainties have been estimated and were kept fixed in the fits; see \autoref{doubly-subst-isos-results}. 
\textsuperscript{\emph{b}} Reference~\citenum{MeCN_rot_2009}. 
\end{table*}

%%%%%%%%%%%%%%%%%%%%%%%%%%%%%%%%%%%%%%%%%%%%%%%%%%%%%%%%%%%%%%%%%%%%%
%%%%%%%%%%%%%%%%%%%%%%%%%%%%%%%%%%%%%%%%%%%%%%%%%%%%%%%%%%%%%%%%%%%%%

The $^{13}$CH$_3$$^{13}$CN isotopolog was already studied quite extensively \cite{MeCN_rot_2009}, albeit in natural isotopic composition. 
These measurements permitted transitions of several $J$ to be identified between 249 and 1139~GHz with $K = 9$ in one transition. 
The present measurements accessed several $J$ between 53 and 927~GHz and reached $K = 16$. 
The previous parameter set \cite{MeCN_rot_2009} was employed to fit the new data; calculations of the initial spectrum were taken from the CDMS. 
It was sufficient to fit $H_J$ in addition to the parameters fit previously to achieve a satisfactory fit. 
The new spectroscopic parameters of $^{13}$CH$_3$$^{13}$CN together with the earlier ones \cite{MeCN_rot_2009} 
and the resulting $^{13}$CH$_3$C$^{15}$N and CH$_3$$^{13}$C$^{15}$N are presented in \autoref{parameter_doubly-subst}.

Spectroscopic parameters of $^{13}$CH$_3$C$^{15}$N and CH$_3$$^{13}$C$^{15}$N were evaluated by assuming $A_0$ to be the same 
for isotopic species with substitutions of C or N atoms and by applying the CH$_3$C$^{15}$N/CH$_3$CN parameter ratios 
and those involving $^{13}$CH$_3$CN or CH$_3$$^{13}$CN to the CH$_3$CN parameters for all others. 
Subsequently, $B$ of each of these doubly substituted species was fit to the data from ref.~\citenum{13C15N-MeCN-rot_1989}, 
which covered $J = 1 - 0$ to $4 - 3$. These earlier data were omitted in the final fits 
because of more accurate data at similar quantum numbers obtained in the course of the present study.

The frequency ranges covered for these isotopomers here are very similar to the ones for $^{13}$CH$_3$$^{13}$CN, 
even more so, as its $B$ value is similar to the CH$_3$$^{13}$C$^{15}$N one, leading to occasional blending of the lines. 
The highest $K$ value in both line lists is 15. 
The parameters fit for these two doubly substituted isotopomers were the same as the ones fit presently for $^{13}$CH$_3$$^{13}$CN. 
The parameter values and their uncertainties, as far as the parameters were fit, are also given in \autoref{parameter_doubly-subst}. 
The line, parameter, and fit files of all three doubly substituted isotopologs are available as Supporting Information 
and have also been deposited in the CDMS as detailed in the Data Availability Statement.

%%%%%%%%%%%%%%%%%%%%%%%%%%%%%%%%%%%%%%%%%%%%%%%%%%%%%%%%%%%%%%%%%%%%%
%%%%%%%%%%%%%%%%%%%%%%%%%%%%%%%%%%%%%%%%%%%%%%%%%%%%%%%%%%%%%%%%%%%%%
\section{DISCUSSION OF THE LABORATORY RESULTS}
\label{lab-discussion}
%%%%%%%%%%%%%%%%%%%%%%%%%%%%%%%%%%%%%%%%%%%%%%%%%%%%%%%%%%%%%%%%%%%%%
%%    Table 5    %%%%%%%%%%%%%%%%%%%%%%%%%%%%%%%%%%%%%%%%%%%%%%%%%%%%
%%%%%%%%%%%%%%%%%%%%%%%%%%%%%%%%%%%%%%%%%%%%%%%%%%%%%%%%%%%%%%%%%%%%%

\begin{table}
\begin{center}
\caption{Comparison of Low Order Spectroscopic Parameters\textsuperscript{\emph{a}} of CH$_3$CN, $^{13}$CH$_3$CN, and CH$_3$$^{13}$CN.}
\label{comparison-table}
% \smallskip
% \renewcommand{\arraystretch}{1.10}
\begin{tabular}[t]{lr@{}lr@{}lr@{}l}
\hline
Parameter       & \multicolumn{2}{c}{CH$_3$CN\textsuperscript{\emph{b}}} & \multicolumn{2}{c}{$^{13}$CH$_3$CN} & \multicolumn{2}{c}{CH$_3$$^{13}$CN} \\
\hline
$(A - B)_0$                     &   148900&.0   &   149165&.6   &   148904&.6   \\
$B_0$                           &     9198&.899 &     8933&.309 &     9194&.350 \\
$eQq$                           &     $-$4&.223 &     $-$4&.218 &     $-$4&.219 \\
$F_2(0,8^{\pm1}) \times 10^3$   &    $-$70&.9   &    $-$56&.3   &         &     \\
$E(8^1)$                        &      365&.024 &      364&.77  &      357&.19\textsuperscript{\emph{c}} \\
$\Delta (A - B)(8^1)$           &   $-$115&.9   &   $-$115&.0\textsuperscript{\emph{d}} &   $-$114&.8\textsuperscript{\emph{d}} \\
$\Delta B(8^1)$                 &       27&.530 &       26&.688 &       26&.405 \\
$A\zeta(8^1)$                   &   138656&.0   &   138859&.4   &   139838&.8   \\
$q(8^1)$                        &       17&.798 &       16&.80  &       18&.21  \\
$\Delta eQq$                    &     $-$0&.039 &     $-$0&.040 &     $-$0&.044 \\
$eQq\eta$                       &        0&.152 &        0&.165 &        0&.156 \\
$F(8^{\pm1},8^{2,\mp2})$        &    53138&.    &    52778&.    &    50585&.    \\
$\Delta E(8^{2^2} - 8^{2^0})$   &       22&.398 &       22&.929 &       21&.792 \\
$\Delta (A - B)(8^{2^0})$       &   $-$187&.5   &   $-$185&.6\textsuperscript{\emph{d}} &   $-$187&.7\textsuperscript{\emph{d}} \\
$\Delta B(8^{2^0})$             &       54&.058 &       52&.100 &       52&.219 \\
$E(8^{2^2})$                    &      739&.148 &      738&.65  &      723&.2   \\
$\Delta (A - B)(8^{2^2})$       &   $-$260&.1   &   $-$260&.7   &   $-$255&.9   \\
$\Delta B(8^{2^2})$             &       54&.503 &       52&.850 &       52&.263 \\
$\Delta A(8^{2^2})$             &   $-$205&.61  &   $-$207&.9   &   $-$203&.7   \\
$A\zeta(8^{2^2})$               &   138655&.4   &   138847&.    &   139828&.    \\
$q(8^2)$                        &       17&.730 &       16&.70  &       18&.18  \\
\hline
\end{tabular}
\end{center}
\textsuperscript{\emph{a}} All parameters in units of MHz, except energies $E$ in units of cm$^{-1}$. 
 Parameters from present fits unless indicated otherwise.  
\textsuperscript{\emph{b}} Reference~\citenum{MeCN_up2v4eq1_etc_2021}. 
\textsuperscript{\emph{c}} From low-resolution IR measurements \cite{FF_Duncan_1978}. 
\textsuperscript{\emph{d}} Derived assuming the isotopic $\Delta A$ values agree with the corresponding CH$_3$CN values. 
\end{table}

%%%%%%%%%%%%%%%%%%%%%%%%%%%%%%%%%%%%%%%%%%%%%%%%%%%%%%%%%%%%%%%%%%%%%
%%%%%%%%%%%%%%%%%%%%%%%%%%%%%%%%%%%%%%%%%%%%%%%%%%%%%%%%%%%%%%%%%%%%%

Perhaps the most interesting results of the present study from the standpoint of molecular physics 
are the precise energy difference determinations between $\varv _8 = 1$ and 2 
of both $^{13}$CH$_3$CN and CH$_3$$^{13}$CN, the difference between $\varv _8 = 0$ and 1 for $^{13}$CH$_3$CN, 
and the differences between $\varv _8 = 2$, $l = 0$ and $\pm2$ for both isotopomers. 
The last type of determination is more generally possible in theory for a degenerate overtone state, but the precision with which this may be achieved 
depends obviously on how close energies get that have $\Delta k = \Delta l = \pm2$ and what the magnitude of $q_{22}$ is.

As can be seen in \autoref{parameter_v8_eq_2}, the $\Delta E(8^{2^2} - 8^{2^0})$ values agree quite well with the initial estimates, slightly less so in the $^{13}$CH$_3$CN case. 
And both values are quite close to the value of the main isotopic species, which are compared with other low-order parameters in \autoref{comparison-table}. 
As mentioned in our previous study on CH$_3$CN \cite{MeCN_up2v4eq1_etc_2021}, this type of energy difference appears to be challenging for quantum chemical calculations. 
Therefore, the present values are additional reference values besides the ones given in ref.~\citenum{MeCN_up2v4eq1_etc_2021} and other values presumably available in the literature. 
Overall, isotopic changes in the low-order parameters in \autoref{comparison-table} are small, and these differences display diverse trends. 
It is gratifying to see that $A\zeta$ and $q$ are very similar in $\varv _8 = 1$ and 2 for each of the $^{13}$C isotopomers, 
even though the values differ somewhat from those of the main isotopic species. 
\autoref{comparison-table} also demonstrates that the HFS parameters $eQq$, $\Delta eQq$, and $eQq_2$ of both isotopomers agree well with the respective values 
$-$4.223, $-$0.039, and 0.152~MHz of the main isotopic species \cite{MeCN_up2v4eq1_etc_2021}. 
These values may also be compared with $-$4.710, $-$0.102, and 0.393~MHz of the lighter HCN molecule \cite{HCN_rot_v2le1_2003}.

Interactions between a degenerate bending state and its overtone or the ground vibrational state have been reported comparatively rarely to the best of our knowledge. 
Examples besides CH$_3$CN are the isoelectronic molecules CH$_3$CCH \cite{MeCCH_nu10+Dyade_2002,CH3CCH_v39_2004,MeCCH_10mue_2009} and CH$_3$NC \cite{MeNC_v8le2_2011}. 
In the case of $^{13}$CH$_3$CN, our $E(8^1)$ value of 364.77~cm$^{-1}$ is 0.21~cm$^{-1}$ higher than the value from low-resolution IR measurements \cite{FF_Duncan_1978}, 
a very good agreement considering that there is no sharp $Q$-branch in a perpendicular $b$-type IR band. 
Our $E(8^{2^0})$ value of 715.73~cm$^{-1}$ is 0.09~cm$^{-1}$ lower than the low-resolution IR value \cite{FF_Duncan_1978}. 
The present $E(8^{2^2})$ value of 738.65~cm$^{-1}$ is about 0.46~cm$^{-1}$ higher than our previous estimate (\autoref{parameter_v8_eq_2}). 
Inspection of the CH$_3$$^{13}$CN values reveals that $E(8^{2^2})$ is nearly 0.2~cm$^{-1}$ higher than our previous estimate, 
and the derived $E(8^{2^0})$ value of 741.40~cm$^{-1}$ is 0.30~cm$^{-1}$ higher than the value from low-resolution IR measurements \cite{FF_Duncan_1978}. 
Since, however, $E(8^{2^2})$ was determined with respect to a fixed $E(8^1)$ value, this may mean the true $E(8^1)$ value is somewhat lower. 
On the other hand, we need to keep in mind that the energy determinations here depend on several parameters that needed to be kept fixed. 
Besides $E(8^1)$ of CH$_3$$^{13}$CN, these are foremost $\Delta (A - B)$ of $\varv _8 = 1$ and $\varv _8 = 2^0$ for both isotopomers 
and possibly also $A - B$ in the ground vibrational state. 
While the last values are challenging to be determined, the $\Delta (A - B)$ values can be obtained from high-resolution IR measurements 
of $\nu _8$ and $2\nu _8$ or the $2\nu _8 - \nu _8$ hot band. 
Furthermore, effects on the spectroscopic parameters may be caused by the neglect of perturbation in $\varv _8 = 2$ by higher vibrational states, 
which may still be non-negligible even though substantial amounts of higher-$k$ data of $l = 0$ and $-2$ were weighted out.

The $\varv _8 = 2^2$ $\Delta A$ values of both isotopomers deviate by $\sim$2~MHz from the value $-$205.61~MHz 
of CH$_3$CN \cite{MeCN_up2v4eq1_etc_2021}, with the $^{13}$CH$_3$CN value being slightly larger in magnitude ($-$207.9~MHz) 
and the CH$_3$$^{13}$CN being slightly smaller ($-$203.7~MHz), see also \autoref{comparison-table}. 
The corresponding $\Delta B$ values agree very well with the previous estimates \cite{MeCN_isos_v8_rot_2016}, 
as can be seen in \autoref{parameter_v8_eq_2}, whereas the $\varv _8 = 2^0$ $\Delta B$ values deviate by slightly more than 0.3~MHz, 
potentially indicating unaccounted perturbations in the $\varv _8 = 2^0$ data included in the fit. 
The $\Delta D_{JK}$ and $\Delta D_{J}$ values point in a similar direction as those of $\varv _8 = 2^2$ are much closer 
to being two times the $\varv _8 = 1$ values than those of $\varv _8 = 2^0$. 
Deviations are more pronounced in the $\Delta H$ values, but still show the $\varv _8 = 2^2$ data to be closer to two times 
the $\varv _8 = 1$ values than those of $\varv _8 = 2^0$. 
Most of the rotational corrections to $A\zeta$ and $q$ were constrained to being the same in $\varv _8 = 1$ and $\varv _8 = 2$; 
unconstrained parameters display various degrees of deviations. 
However, we should keep in mind that even in the latest account on the main isotopic species \cite{MeCN_up2v4eq1_etc_2021} 
some parameters remain which do not exhibit the usual $\varv _8 = 1$ to $\varv _8 = 2$ ratios. 
The $F(8^{\pm1},8^{2,\mp2})$ values are now quite well determined, showing some deviations from the previous values. 
The $^{13}$CH$_3$CN value is now only slightly smaller than the 53.14~GHz of CH$_3$CN \cite{MeCN_up2v4eq1_etc_2021}, 
whereas the CH$_3$$^{13}$CN value was previously somewhat larger than the CH$_3$CN value and now considerably smaller. 
However, these values may change somewhat upon better accounting of the higher-$K$ data of the $l = 0$ and $-2$ substates.

Little needs to be mentioned concerning the present $\varv _8 = 1$ and $\varv = 0$ values in comparison to previous data. 
The uncertainties improved, in some cases around a factor of 10, because of the increased data sets even though additional parameters were floated in the fits. 
Changes in the lower order parameters are small, as it should be, while changes in some higher order parameters are not so small, which is not so unusual either. 
The interaction parameter $F_2(0,8^{\pm1})$ of $^{13}$CH$_3$CN is $-$56.3~kHz compared to $-$70.9~kHz for CH$_3$CN. 
The value of the $^{13}$C species may well be affected by the chosen values of $\Delta (A - B)$ and other fixed parameters. 
We also obtained greatly improved spectroscopic parameters for $^{13}$CH$_3$$^{13}$CN along with extensive parameter sets for $^{13}$CH$_3$C$^{15}$N and CH$_3$$^{13}$C$^{15}$N.

%%%%%%%%%%%%%%%%%%%%%%%%%%%%%%%%%%%%%%%%%%%%%%%%%%%%%%%%%%%%%%%%%%%%%
%%%%%%%%%%%%%%%%%%%%%%%%%%%%%%%%%%%%%%%%%%%%%%%%%%%%%%%%%%%%%%%%%%%%%
\section{ASTRONOMICAL RESULTS}
\label{astro}
%%%%%%%%%%%%%%%%%%%%%%%%%%%%%%%%%%%%%%%%%%%%%%%%%%%%%%%%%%%%%%%%%%%%%
%%%%%%%%%%%%%%%%%%%%%%%%%%%%%%%%%%%%%%%%%%%%%%%%%%%%%%%%%%%%%%%%%%%%%

%%%%%%%%%%%%%%%%%%%%%%%%%%%%%%%%%%%%%%%%%%%%%%%%%%%%%%%%%%%%%%%%%%%%%

\begin{figure*}
\centerline{\resizebox{0.85\hsize}{!}{\includegraphics[angle=0]{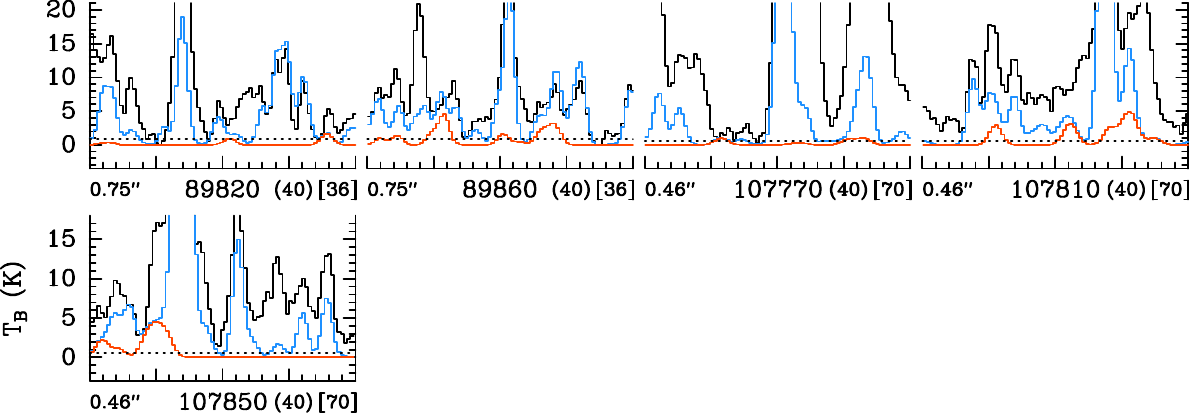}}}
\caption{Transitions of $^{13}$CH$_3$CN $\varv_8 = 2$ covered by the ReMoCA survey. 
The LTE synthetic spectrum of $^{13}$CH$_3$CN $\varv_8 = 2$ is displayed in red and overlaid on the spectrum observed toward Sgr~B2(N1S) shown in black. 
The blue synthetic spectrum contains the contributions of all molecules identified in our survey so far, including $^{13}$CH$_3$CN $\varv_8 = 2$. 
The values written below each panel correspond from left to right to the half-power beam width, the central frequency in MHz, 
the width in MHz of each panel in parentheses, and the continuum level in K of the baseline-subtracted spectra in brackets. 
The $y$-axis is labeled in brightness temperature units (K). The dotted line indicates the $3\sigma$ noise level.}
\label{f:remoca_13c1}
\end{figure*}

%%%%%%%%%%%%%%%%%%%%%%%%%%%%%%%%%%%%%%%%%%%%%%%%%%%%%%%%%%%%%%%%%%%%%

\begin{figure*}
\centerline{\resizebox{0.85\hsize}{!}{\includegraphics[angle=0]{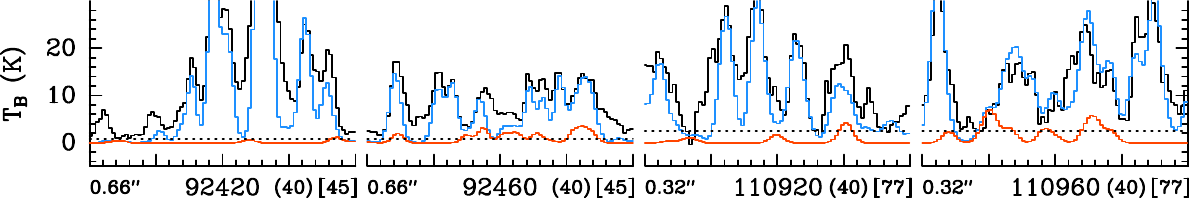}}}
\caption{Same as \autoref{f:remoca_13c1}, but for CH$_3$$^{13}$CN $\varv_8 = 2$.}
\label{f:remoca_13c2}
\end{figure*}

%%%%%%%%%%%%%%%%%%%%%%%%%%%%%%%%%%%%%%%%%%%%%%%%%%%%%%%%%%%%%%%%%%%%%
%%%%%%%%%%%%%%%%%%%%%%%%%%%%%%%%%%%%%%%%%%%%%%%%%%%%%%%%%%%%%%%%%%%%%

We used the spectroscopic results obtained in \autoref{spec-fitting} to search for rotational emission 
from within the $\varv_8 = 2$ vibrational state of the $^{13}$C isotopomers of methyl cyanide 
toward the main hot molecular core of the Sgr~B2(N) star-forming region. 
We employed the imaging spectral line survey ReMoCA carried out with ALMA between 84 and 114~GHz. 
Details about the observations and method of analysis of the survey can be found in Belloche et al. \cite{ReMoCA_Sgr-B2(N2)_2025}. 
The spectra were analyzed under the assumption of local thermodynamic equilibrium with the Weeds software \cite{Weeds_2011}.

M{\"u}ller et al. \cite{MeCN_up2v4eq1_etc_2021} presented a detailed analysis of the emission of methyl cyanide 
in its vibrational ground state and several of its vibrationally excited states as well as its $^{13}$C isotopologs 
in their vibrational ground state and in $\varv_8=1$, toward the position called Sgr~B2(N1S), which is located within 
the main hot molecular core of Sgr~B2(N). Using the same LTE parameters as in ref.~\citenum{MeCN_up2v4eq1_etc_2021}, 
that is a column density of $1.4 \times 10^{17}$~cm$^{-2}$, a temperature of 260~K, a line width of 6~km~s$^{-1}$, 
and a systemic velocity of 61.8~km~s$^{-1}$, we computed synthetic spectra for the $^{13}$C isotopologs in $\varv_8=2$. 
This column density corresponds to a $^{12}$C/$^{13}$C ratio of 21, which is typical for the Galactic center 
region\cite{13C-VyCN_2008,EMoCA_2016,ROH_RSH_EMoCA_2016,12C-13C_SgrB2N_2017,det-13CH_2020}.

The synthetic spectra of $^{13}$CH$_3$CN $\varv_8=2$ and CH$_3$$^{13}$CN $\varv_8=2$ are displayed in red 
and compared to the ReMoCA spectra shown in black in \autoref{f:remoca_13c1} and \autoref{f:remoca_13c2}, respectively. 
The contributions of all molecules identified so far toward Sgr~B2(N1S), including $^{13}$CH$_3$CN $\varv_8=2$ 
and CH$_3$$^{13}$CN $\varv_8=2$, is overlaid in blue. 
The synthetic spectra are consistent with the observed spectra with several lines above the $3\sigma$ noise limit, 
but no line of $^{13}$CH$_3$CN $\varv_8=2$ and CH$_3$$^{13}$CN $\varv_8=2$ is sufficiently free of contamination 
from other species to allow us to claim a robust identification of these states.

%%%%%%%%%%%%%%%%%%%%%%%%%%%%%%%%%%%%%%%%%%%%%%%%%%%%%%%%%%%%%%%%%%%%%
%%%%%%%%%%%%%%%%%%%%%%%%%%%%%%%%%%%%%%%%%%%%%%%%%%%%%%%%%%%%%%%%%%%%%
\section{CONCLUSION}
\label{conclusion}
%%%%%%%%%%%%%%%%%%%%%%%%%%%%%%%%%%%%%%%%%%%%%%%%%%%%%%%%%%%%%%%%%%%%%
%%%%%%%%%%%%%%%%%%%%%%%%%%%%%%%%%%%%%%%%%%%%%%%%%%%%%%%%%%%%%%%%%%%%%

We employed methyl cyanide samples enriched in $^{13}$CH$_3$CN and CH$_3$$^{13}$CN to study their 
$\varv _8 = 2$ excited states extensively and to extend $\varv _8 = 1$ and 0 data sets. 
Perturbations sampled the energy differences of the $l = 0$ and $l = \pm2$ substates of $\varv _8 = 2$, 
the energy of $\varv _8 = 2$, and in the case of $^{13}$CH$_3$CN the energy of $\varv _8 = 1$, 
thus improving our knowledge on the molecular properties of methyl cyanide.

A search for lines pertaining to $\varv _8 = 2$ of $^{13}$CH$_3$CN and CH$_3$$^{13}$CN toward Sgr~B2(N1S) 
suggests that some of these lines are above the $3\sigma$ limit, but contamination by other species prevents a secure identification. 
Nevertheless, the results indicate that transitions of such high vibrational states and in isotopic species 
may well be identifiable in warm and dense parts of star-forming regions.

We also improved the $^{13}$CH$_3$$^{13}$CN data set and obtained extensive spectroscopic parameters of 
$^{13}$CH$_3$C$^{15}$N and CH$_3$$^{13}$C$^{15}$N that are accurate enough to search for these rare isotopomers in space, 
even though the chances of finding them may be quite small.

%%%%%%%%%%%%%%%%%%%%%%%%%%%%%%%%%%%%%%%%%%%%%%%%%%%%%%%%%%%%%%%%%%%%%
%%%%%%%%%%%%%%%%%%%%%%%%%%%%%%%%%%%%%%%%%%%%%%%%%%%%%%%%%%%%%%%%%%%%%

\section*{ACKNOWLEDGMENTS}

We acknowledge support by the Deutsche Forschungsgemeinschaft (DFG) via the collaborative research center 
SFB~1601 (Project 500700252) subprojects A4 and Inf as well as the Ger{\"a}tezentrum SCHL~341/15-1 (``Cologne Center for Terahertz Spectroscopy''). 
We thank Brian J. Drouin and John C. Pearson from JPL for the methyl cyanide spectral recordings taken almost 20 years ago. 
We thank the Regionales Rechenzentrum der Universit{\"a}t zu K{\"o}ln for computing time.
Our research benefited from NASA's Astrophysics Data System. 
This paper makes use of the following ALMA data: ADS/JAO.ALMA \# 2016.1.00074.S. 
ALMA is a partnership of ESO (representing its member states), NSF (USA) and NINS (Japan), 
together with NRC (Canada), MOST and ASIAA (Taiwan), and KASI (Republic of Korea), 
in cooperation with the Republic of Chile. The Joint ALMA Observatory is operated by ESO, 
AUI/NRAO and NAOJ.

%%%%%%%%%%%%%%%%%%%%%%%%%%%%%%%%%%%%%%%%%%%%%%%%%%%%%%%%%%%%%%%%%%%%%
%%%%%%%%%%%%%%%%%%%%%%%%%%%%%%%%%%%%%%%%%%%%%%%%%%%%%%%%%%%%%%%%%%%%%

\section*{Data Availability Statement}
The input and output files of SPFIT are provided as supplementary material. 
These files as well as auxiliary files have been deposited in the data section of the CDMS \cite{CDMS-data}. 
Calculations of rotational spectra will be available in the catalog section of the CDMS \cite{CDMS-entries}.


\section{Supporting Information}

%A listing of the contents of each file supplied as Supporting Information
%should be included. For instructions on what should be included in the
%Supporting Information as well as how to prepare this material for
%publications, refer to the journal's Instructions for Authors.

The Supporting Information is available free of charge at 
\href{https://pubs.acs.org/doi/10.1021/acsearthspace-chem.5c00353}{https://pubs.acs.org/doi/10.1021/acsearthspace-chem.5c00353.}

The line, parameter, and fit files (with extensions .lin, .par, and .fit) 
are provided for all five isotopic species. The first file, numbered 001, explains the connection of the numbers 
with the respective isotopolog. All files are text files.

\bibliographystyle{elsarticle-num}
\bibliography{MeCN}

@ARTICLE{MeCN_det_T_1971,
       author = {{Solomon}, P.~M. and {Jefferts}, K.~B. and {Penzias}, A.~A. and
         {Wilson}, R.~W.},
        title = "{Detection of Millimeter Emission Lines from Interstellar Methyl Cyanide}",
      journal = {Astrophys. J. Lett.},
         year = "1971",
        month = "Sep",
       volume = {168},
        pages = {L107-L110},
          doi = {10.1086/180794},
       adsurl = {https://ui.adsabs.harvard.edu/abs/1971ApJ...168L.107S}
}

@misc{astrochymist,
note = {See for example the Astrochymist Interstellar \& Circumstellar Molecules page at http://www.astrochymist.org/astrochymist\_ism.html; accessed 2025-11-10.}
}

@ARTICLE{MeCN_IRAS_16293_2003,
       author = {{Cazaux}, S. and {Tielens}, A.~G.~G.~M. and {Ceccarelli}, C. and {Castets}, A. and {Wakelam}, V. and {Caux}, E. and {Parise}, B. and {Teyssier}, D.},
        title = "{The Hot Core around the Low-mass Protostar IRAS 16293-2422: Scoundrels Rule!}",
      journal = {Astrophys. J.},
         year = 2003,
        month = aug,
       volume = {593},
       number = {1},
        pages = {L51-L55},
          doi = {10.1086/378038},
       adsurl = {https://ui.adsabs.harvard.edu/abs/2003ApJ...593L..51C}
}

@ARTICLE{MeCN-TMC-1_1983,
       author = {{Matthews}, H.~E. and {Sears}, T.~J.},
        title = "{Detection of the $J = 1 \to 0$ transition of CH$_3$CN.}",
      journal = {Astrophys. J.},
         year = 1983,
        month = apr,
       volume = {267},
        pages = {L53-L57},
          doi = {10.1086/184001},
       adsurl = {https://ui.adsabs.harvard.edu/abs/1983ApJ...267L..53M}
}

@ARTICLE{MeCN-CSE_1984,
       author = {{Johansson}, L.~E.~B. and {Andersson}, C. and {Ellder}, J. and {Friberg}, P. and {Hjalmarson}, A. and {Hoglund}, B. and {Irvine}, W.~M. and {Olofsson}, H. and {Rydbeck}, G.},
        title = "{Spectral scan of Orion A and IRC +10216 from 72 to 91 GHz.}",
      journal = {Astron. Astrophys.},
         year = 1984,
        month = jan,
       volume = {130},
        pages = {227-256},
       adsurl = {https://ui.adsabs.harvard.edu/abs/1984A&A...130..227J}
}

@ARTICLE{MeCN_MeCCH_extragal,
       author = {{Mauersberger}, R. and {Henkel}, C. and {Walmsley}, C.~M. and {Sage}, L.~J. and {Wiklind}, T.},
        title = "{Dense gas in nearby galaxies. V. Multilevel studies of CH$_3$CCH and CH$_3$CN.}",
      journal = {Astron. Astrophys.},
         year = 1991,
        month = jul,
       volume = {247},
        pages = {307},
       adsurl = {https://ui.adsabs.harvard.edu/abs/1991A&A...247..307M}
}

@ARTICLE{MeCN_PP-disk_2015,
       author = {{{\"O}berg}, Karin I. and {Guzm{\'a}n}, Viviana V. and {Furuya}, Kenji and {Qi}, Chunhua and {Aikawa}, Yuri and {Andrews}, Sean M. and {Loomis}, Ryan and {Wilner}, David J.},
        title = "{The comet-like composition of a protoplanetary disk as revealed by complex cyanides}",
      journal = {Nature},
         year = 2015,
        month = apr,
       volume = {520},
       number = {7546},
        pages = {198-201},
          doi = {10.1038/nature14276},
archivePrefix = {arXiv},
       eprint = {1505.06347},
 primaryClass = {astro-ph.GA},
       adsurl = {https://ui.adsabs.harvard.edu/abs/2015Natur.520..198O}
}

@ARTICLE{translucent-toward-SgrB2_2019,
       author = {{Thiel}, V. and {Belloche}, A. and {Menten}, K.~M. and {Giannetti}, A. and {Wiesemeyer}, H. and {Winkel}, B. and {Gratier}, P. and {M{\"u}ller}, H.~S.~P. and {Colombo}, D. and {Garrod}, R.~T.},
        title = "{Small-scale physical and chemical structure of diffuse and translucent molecular clouds along the line of sight to Sgr B2}",
      journal = {Astron. Astrophys.},
         year = 2019,
        month = mar,
       volume = {623},
          eid = {A68},
        pages = {A68},
          doi = {10.1051/0004-6361/201834467},
archivePrefix = {arXiv},
       eprint = {1901.03231},
 primaryClass = {astro-ph.GA},
       adsurl = {https://ui.adsabs.harvard.edu/abs/2019A&A...623A..68T}
}

@ARTICLE{biomass_burning_MLS_2004,
       author = {{Livesey}, Nathaniel J. and {Fromm}, Michael D. and {Waters}, Joe W. and {Manney}, Gloria L. and {Santee}, Michelle L. and {Read}, William G.},
        title = "{Enhancements in lower stratospheric CH$_{3}$CN observed by the Upper Atmosphere Research Satellite Microwave Limb Sounder following boreal forest fires}",
      journal = {J. Geophys. Res.},
         year = 2004,
        month = mar,
       volume = {109},
       number = {D6},
          eid = {D06308},
        pages = {D06308},
          doi = {10.1029/2003JD004055},
       adsurl = {https://ui.adsabs.harvard.edu/abs/2004JGRD..109.6308L}
}

@ARTICLE{biomass_burning_2011,
       author = {{Simpson}, I.~J. and {Akagi}, S.~K. and {Barletta}, B. and {Blake}, N.~J. and {Choi}, Y. and {Diskin}, G.~S. and {Fried}, A. and {Fuelberg}, H.~E. and {Meinardi}, S. and {Rowland}, F.~S. and {Vay}, S.~A. and {Weinheimer}, A.~J. and {Wennberg}, P.~O. and {Wiebring}, P. and {Wisthaler}, A. and {Yang}, M. and {Yokelson}, R.~J. and {Blake}, D.~R.},
        title = "{Boreal forest fire emissions in fresh Canadian smoke plumes: C$_{1}$-C$_{10}$ volatile organic compounds (VOCs), CO$_{2}$, CO, NO$_{2}$, NO, HCN and CH$_{3}$CN}",
      journal = {Atmos. Chem. Phys.},
         year = 2011,
        month = jul,
       volume = {11},
       number = {13},
        pages = {6445-6463},
          doi = {10.5194/acp-11-6445-201110.5194/acpd-11-9515-2011},
       adsurl = {https://ui.adsabs.harvard.edu/abs/2011ACP....11.6445S}
}

@ARTICLE{MeCN-Kohoutek_1974,
       author = {{Ulich}, B.~L. and {Conklin}, E.~K.},
        title = "{Detection of methyl cyanide in Comet Kohoutek}",
      journal = {Nature},
         year = 1974,
        month = mar,
       volume = {248},
       number = {5444},
        pages = {121-122},
          doi = {10.1038/248121a0},
       adsurl = {https://ui.adsabs.harvard.edu/abs/1974Natur.248..121U}
}

@INPROCEEDINGS{MeCN-Titan_1993,
       author = {{B{\'e}zard}, B. and {Marten}, A. and {Paubert}, G.},
        title = "{Detection of Acetonitrile on Titan}",
    booktitle = {AAS/Division for Planetary Sciences Meeting Abstracts \#25},
         year = 1993,
       series = {AAS/Division for Planetary Sciences Meeting Abstracts},
       volume = {25},
        month = jun,
          eid = {25.09},
        pages = {25.09},
       adsurl = {https://ui.adsabs.harvard.edu/abs/1993DPS....25.2509B}
}

@ARTICLE{CH3CN_v8eq1_det_1983,
       author = {{Goldsmith}, P.~F. and {Krotkov}, R. and {Snell}, R.~L. and {Brown}, R.~D. and {Godfrey}, P.},
        title = "{Vibrationally excited CH$_3$CN and HC$_3$N in Orion.}",
      journal = {Astrophys. J.},
         year = 1983,
        month = nov,
       volume = {274},
        pages = {184-194},
          doi = {10.1086/161436},
       adsurl = {https://ui.adsabs.harvard.edu/abs/1983ApJ...274..184G}
}

@ARTICLE{OrionKL-ALMA_CH3CNv8eq2_2012,
       author = {{Fortman}, Sarah M. and {McMillan}, James P. and {Neese}, Christopher F. and {Randall}, Suzanna K. and {Remijan}, Anthony J. and {Wilson}, T.~L. and {De Lucia}, Frank C.},
        title = "{An analysis of a preliminary ALMA Orion KL spectrum via the use of complete experimental spectra from the laboratory}",
      journal = {J. Mol. Spectrosc.},
         year = 2012,
        month = oct,
       volume = {280},
        pages = {11-20},
          doi = {10.1016/j.jms.2012.08.002},
       adsurl = {https://ui.adsabs.harvard.edu/abs/2012JMoSp.280...11F}
}

@ARTICLE{SgrB2-survey_2013,
       author = {{Belloche}, A. and {M{\"u}ller}, H.~S.~P. and {Menten}, K.~M. and {Schilke}, P. and {Comito}, C.},
        title = "{Complex organic molecules in the interstellar medium: IRAM 30 m line survey of Sagittarius B2(N) and (M)}",
      journal = {Astron. Astrophys.},
         year = 2013,
        month = nov,
       volume = {559},
          eid = {A47},
        pages = {A47},
          doi = {10.1051/0004-6361/201321096},
archivePrefix = {arXiv},
       eprint = {1308.5062},
 primaryClass = {astro-ph.GA},
       adsurl = {https://ui.adsabs.harvard.edu/abs/2013A&A...559A..47B}
}

@ARTICLE{MeCN_up2v4eq1_etc_2021,
       author = {{M{\"u}ller}, Holger S.~P. and {Belloche}, Arnaud and {Lewen}, Frank and {Drouin}, Brian J. and {Sung}, Keeyoon and {Garrod}, Robin T. and {Menten}, Karl M.},
        title = "{Toward a global model of the interactions in low-lying states of methyl cyanide: Rotational and rovibrational spectroscopy of the $\varv _{4} = 1$ state and tentative interstellar detection of the $\varv _{4} = \varv _{8} = 1$ state in Sgr B2(N)}",
      journal = {J. Mol. Spectrosc.},
         year = 2021,
        month = apr,
       volume = {378},
          eid = {111449},
        pages = {111449},
          doi = {10.1016/j.jms.2021.111449},
archivePrefix = {arXiv},
       eprint = {2103.07389},
 primaryClass = {astro-ph.GA},
       adsurl = {https://ui.adsabs.harvard.edu/abs/2021JMoSp.37811449M}
}

@ARTICLE{CH3CN_w_CH313CN_SgrB2_1983,
       author = {{Cummins}, S.~E. and {Green}, S. and {Thaddeus}, P. and {Linke}, R.~A.},
        title = "{The kinetic temperature and density of the Sagittarius B2 molecular cloud from observations of methyl cyanide.}",
      journal = {Astrophys. J.},
         year = 1983,
        month = mar,
       volume = {266},
        pages = {331-338},
          doi = {10.1086/160782},
       adsurl = {https://ui.adsabs.harvard.edu/abs/1983ApJ...266..331C}
}

@ARTICLE{Orion-A_survey_1985,
       author = {{Sutton}, E.~C. and {Blake}, G.~A. and {Masson}, C.~R. and {Phillips}, T.~G.},
        title = "{Molecular line survey of Orion A from 215 to 247 GHz.}",
      journal = {Astrophys. J. Suppl. Ser.},
         year = 1985,
        month = jul,
       volume = {58},
        pages = {341-378},
          doi = {10.1086/191045},
       adsurl = {https://ui.adsabs.harvard.edu/abs/1985ApJS...58..341S}
}

@ARTICLE{det_CH2DCN_1992,
       author = {{Gerin}, M. and {Combes}, F. and {Wlodarczak}, G. and {Jacq}, T. and {Guelin}, M. and {Encrenaz}, P. and {Laurent}, C.},
        title = "{Interstellar detection of deuterated methyl cyanide.}",
      journal = {Astron. Astrophys.},
         year = 1992,
        month = jun,
       volume = {259},
        pages = {L35-L38},
       adsurl = {https://ui.adsabs.harvard.edu/abs/1992A&A...259L..35G}
}

@ARTICLE{SgrB2_survey_1998,
       author = {{Nummelin}, A. and {Bergman}, P. and {Hjalmarson}, {\r{A}}. and {Friberg}, P. and {Irvine}, W.~M. and {Millar}, T.~J. and {Ohishi}, M. and {Saito}, S.},
        title = "{A Three-Position Spectral Line Survey of Sagittarius B2 between 218 and 263 GHz. I. The Observational Data}",
      journal = {Astrophys. J. Suppl. Ser.},
         year = 1998,
        month = jul,
       volume = {117},
       number = {2},
        pages = {427-529},
          doi = {10.1086/313126},
       adsurl = {https://ui.adsabs.harvard.edu/abs/1998ApJS..117..427N}
}

@ARTICLE{EMoCA_2016,
       author = {{Belloche}, A. and {M{\"u}ller}, H.~S.~P. and {Garrod}, R.~T. and {Menten}, K.~M.},
        title = "{Exploring molecular complexity with ALMA (EMoCA): Deuterated complex organic molecules in Sagittarius B2(N2)}",
      journal = {Astron. Astrophys.},
         year = "2016",
        month = "Mar",
       volume = {587},
          eid = {A91},
        pages = {A91},
          doi = {10.1051/0004-6361/201527268},
archivePrefix = {arXiv},
       eprint = {1511.05721},
 primaryClass = {astro-ph.GA},
       adsurl = {https://ui.adsabs.harvard.edu/abs/2016A&A...587A..91B}
}

@ARTICLE{CHD2CN_det_2018,
       author = {{Calcutt}, H. and {J{\o}rgensen}, J.~K. and {M{\"u}ller}, H.~S.~P. and {Kristensen}, L.~E. and {Coutens}, A. and {Bourke}, T.~L. and {Garrod}, R.~T. and {Persson}, M.~V. and {van der Wiel}, M.~H.~D. and {van Dishoeck}, E.~F. and {Wampfler}, S.~F.},
        title = "{The ALMA-PILS survey: complex nitriles towards IRAS 16293-2422}",
      journal = {Astron. Astrophys.},
         year = 2018,
        month = aug,
       volume = {616},
          eid = {A90},
        pages = {A90},
          doi = {10.1051/0004-6361/201732289},
archivePrefix = {arXiv},
       eprint = {1804.09210},
 primaryClass = {astro-ph.GA},
       adsurl = {https://ui.adsabs.harvard.edu/abs/2018A&A...616A..90C}
}

@ARTICLE{MeCN_rot_2009,
       author = {{M{\"u}ller}, H.~S.~P. and {Drouin}, B.~J. and {Pearson}, J.~C.},
        title = "{Rotational spectra of isotopic species of methyl cyanide, CH$_3$CN, in their ground vibrational states up to terahertz frequencies}",
      journal = {Astron. Astrophys.},
         year = 2009,
        month = nov,
       volume = {506},
       number = {3},
        pages = {1487-1499},
          doi = {10.1051/0004-6361/200912932},
archivePrefix = {arXiv},
       eprint = {0910.3111},
 primaryClass = {astro-ph.GA},
       adsurl = {https://ui.adsabs.harvard.edu/abs/2009A&A...506.1487M}
}

@ARTICLE{MeCN_v8le2_2015,
       author = {{M{\"u}ller}, Holger S.~P. and {Brown}, Linda R. and {Drouin}, Brian J. and
         {Pearson}, John C. and {Kleiner}, Isabelle and {Sams}, Robert L. and
         {Sung}, Keeyoon and {Ordu}, Matthias H. and {Lewen}, Frank},
        title = "{Rotational spectroscopy as a tool to investigate interactions between vibrational polyads in symmetric top molecules: Low-lying states $\varv_{8} \leq 2$ of methyl cyanide, CH$_{3}$CN}",
      journal = {J. Mol. Spectrosc.},
         year = "2015",
        month = "Jun",
       volume = {312},
        pages = {22-37},
          doi = {10.1016/j.jms.2015.02.009},
archivePrefix = {arXiv},
       eprint = {1502.06867},
 primaryClass = {astro-ph.GA},
       adsurl = {https://ui.adsabs.harvard.edu/abs/2015JMoSp.312...22M}
}

@ARTICLE{MeCN_isos_v8_rot_2016,
       author = {{M{\"u}ller}, Holger S.~P. and {Drouin}, Brian J. and
         {Pearson}, John C. and {Ordu}, Matthias H. and {Wehres}, Nadine and
         {Lewen}, Frank},
        title = "{Rotational spectra of isotopic species of methyl cyanide, CH$_{3}$CN, in their $\varv_{8} = 1$ excited vibrational states}",
      journal = {Astron. Astrophys.},
         year = "2016",
        month = "Feb",
       volume = {586},
          eid = {A17},
        pages = {A17},
          doi = {10.1051/0004-6361/201527602},
archivePrefix = {arXiv},
       eprint = {1512.05271},
 primaryClass = {astro-ph.GA},
       adsurl = {https://ui.adsabs.harvard.edu/abs/2016A&A...586A..17M}
}

@ARTICLE{13C15N-MeCN-rot_1989,
       author = {{Tam}, H.~S. and {Roberts}, J.~A.},
        title = "{The vibration-rotation microwave spectrum of $^{13}$C tagged acetonitrile in the region 17 to 75~GHz for the ground, $\varv _{8} = 1$ and 2 vibrational states}",
      journal = {J. Mol. Spectrosc.},
         year = 1989,
        month = apr,
       volume = {134},
       number = {2},
        pages = {281-289},
          doi = {10.1016/0022-2852(89)90314-7},
       adsurl = {https://ui.adsabs.harvard.edu/abs/1989JMoSp.134..281T}
}

@ARTICLE{n-BuCN_rot_2012,
       author = {{Ordu}, M.~H. and {M{\"u}ller}, H.~S.~P. and {Walters}, A. and
         {Nu{\~n}ez}, M. and {Lewen}, F. and {Belloche}, A. and {Menten}, K.~M. and
         {Schlemmer}, S.},
        title = "{The quest for complex molecules in space: laboratory spectroscopy of n-butyl cyanide, n-C$_{4}$H$_{9}$CN, in the millimeter wave region and its astronomical search in Sagittarius B2(N)}",
      journal = {Astron. Astrophys.},
         year = 2012,
        month = may,
       volume = {541},
          eid = {A121},
        pages = {A121},
          doi = {10.1051/0004-6361/201118738},
archivePrefix = {arXiv},
       eprint = {1204.2686},
 primaryClass = {astro-ph.GA},
       adsurl = {https://ui.adsabs.harvard.edu/abs/2012A&A...541A.121O}
}

@ARTICLE{OSSO_rot_2015,
       author = {{Martin-Drumel}, M.~A. and {van Wijngaarden}, J. and {Zingsheim}, O. and 
         {Lewen}, F. and {Harding}, M.~E. and {Schlemmer}, S. and {Thorwirth}, S.},
        title = "{Millimeter- and submillimeter-wave spectroscopy of disulfur dioxide, OSSO}",
      journal = {J. Mol. Spectrosc.},
         year = 2015,
        month = jan,
       volume = {307},
        pages = {33-39},
          doi = {10.1016/j.jms.2014.11.007},
       adsurl = {https://ui.adsabs.harvard.edu/abs/2015JMoSp.307...33M}
}

@ARTICLE{MeSH_rot_2012,
       author = {{Xu}, Li-Hong and {Lees}, R.~M. and {Crabbe}, G.~T. and
         {Myshrall}, J.~A. and {M{\"u}ller}, H.~S.~P. and {Endres}, C.~P. and
         {Baum}, O. and {Lewen}, F. and {Schlemmer}, S. and {Menten}, K.~M. and
         {Billinghurst}, B.~E.},
        title = "{Terahertz and far-infrared synchrotron spectroscopy and global modeling of methyl mercaptan, CH$_{3}$$^{32}$SH}",
      journal = {J. Chem. Phys.},
         year = 2012,
        month = sep,
       volume = {137},
       number = {10},
          eid = {104313},
        pages = {104313},
          doi = {10.1063/1.4745792},
       adsurl = {https://ui.adsabs.harvard.edu/abs/2012JChPh.137j4313X}
}

@ARTICLE{JPL_multiplier_spectrometer_2005,
       author = {{Drouin}, Brian J. and {Maiwald}, Frank W. and {Pearson}, John C.},
        title = "{Application of cascaded frequency multiplication to molecular spectroscopy}",
      journal = {Rev. Sci. Instr.},
         year = 2005,
        month = sep,
       volume = {76},
       number = {9},
          eid = {093113},
        pages = {093113},
          doi = {10.1063/1.2042687},
       adsurl = {https://ui.adsabs.harvard.edu/abs/2005RScI...76i3113D}
}

@ARTICLE{MeCN-dipole,
       author = {{Gadhi}, J. and {Lahrouni}, A. and {Legrand}, J. and {Demaison}, J.},
        title = "{Dipole moment of CH$_3$CN}",
      journal = {J. Chim. Phys. Phys.-Chim. Biol.},
         year = 1995,
       volume = {92},
        pages = {1984-1992},
          doi = {10.1051/jcp/1995921984}
}

@ARTICLE{MeCN_DeltaK=3_1993,
       author = {{Anttila}, R. and {Horneman}, V.~M. and {Koivusaari}, M. and {Paso}, R.},
        title = "{Ground State Constants $A_{0}$, $D^{K}_{0}$ and $H^{K}_{0}$ of CH$_{3}$CN}",
      journal = {J. Mol. Spectrosc.},
         year = 1993,
        month = jan,
       volume = {157},
       number = {1},
        pages = {198-207},
          doi = {10.1006/jmsp.1993.1016},
       adsurl = {https://ui.adsabs.harvard.edu/abs/1993JMoSp.157..198A}
}

@ARTICLE{Becke_1993,
       author = {{Becke}, Axel D.},
        title = "{Density-functional thermochemistry. III. The role of exact exchange}",
      journal = {J. Chem. Phys.},
         year = 1993,
        month = apr,
       volume = {98},
       number = {7},
        pages = {5648-5652},
          doi = {10.1063/1.464913},
       adsurl = {https://ui.adsabs.harvard.edu/abs/1993JChPh..98.5648B}
}

@ARTICLE{LYP_1988,
       author = {{Lee}, Chengteh and {Yang}, Weitao and {Parr}, Robert G.},
        title = "{Development of the Colle-Salvetti correlation-energy formula into a functional of the electron density}",
      journal = {Phys. Rev. B},
         year = 1988,
        month = jan,
       volume = {37},
       number = {2},
        pages = {785-789},
          doi = {10.1103/PhysRevB.37.785},
       adsurl = {https://ui.adsabs.harvard.edu/abs/1988PhRvB..37..785L}
}

@ARTICLE{cc-pVXZ_1989,
       author = {{Dunning, Jr.}, Thom H.},
        title = "{Gaussian basis sets for use in correlated molecular calculations. I. The atoms boron through neon and hydrogen}",
      journal = {J. Chem. Phys.},
         year = "1989",
        month = "Jan",
       volume = {90},
       number = {2},
        pages = {1007-1023},
          doi = {10.1063/1.456153},
       adsurl = {https://ui.adsabs.harvard.edu/abs/1989JChPh..90.1007D}
}

@misc{Gaussian03B,
author = {{Frisch}, M.~J. and {Trucks}, G.~W. and {Schlegel}, H.~B. and {Scuseria}, G.~E. and {Robb}, M.~A. and 
          {Cheeseman}, J.~R. and {Montgomery, Jr.}, J.~A. and {Vreven}, T. and {Kudin}, K.~N. and {et al.}},
note = {{Gaussian~03}, Revision B.04, Gaussian, Inc., Wallingford CT}, 
year = 2003
}

@ARTICLE{spfit_1991,
       author = {{Pickett}, Herbert M.},
        title = "{The fitting and prediction of vibration-rotation spectra with spin interactions}",
      journal = {J. Mol. Spectrosc.},
         year = "1991",
        month = "Aug",
       volume = {148},
       number = {2},
        pages = {371-377},
          doi = {10.1016/0022-2852(91)90393-O},
       adsurl = {https://ui.adsabs.harvard.edu/abs/1991JMoSp.148..371P}
}

@ARTICLE{spins-in-spfit_2004,
       author = {{Pickett}, Herbert M.},
        title = "{Spin eigenfunctions and operators for the $D_{\rm n}$ groups}",
      journal = {J. Mol. Spectrosc.},
         year = 2004,
        month = dec,
       volume = {228},
       number = {2},
        pages = {659-663},
          doi = {10.1016/j.jms.2004.05.012},
       adsurl = {https://ui.adsabs.harvard.edu/abs/2004JMoSp.228..659P}
}

@ARTICLE{editorial_Herb-Ed,
       author = {{Drouin}, Brian J. and {M{\"u}ller}, Holger S. P.},
        title = "{Special issue dedicated to the pioneering work of Drs. Edward A. Cohen and Herbert M. Pickett on spectroscopy relevant to the Earth's atmosphere and astrophysics}",
      journal = {J. Mol. Spectrosc.},
         year = 2008,
        month = sep,
       volume = {251},
       number = {1-2},
        pages = {1-3},
          doi = {10.1016/j.jms.2008.05.004},
       adsurl = {https://ui.adsabs.harvard.edu/abs/2008JMoSp.251....1D}
}

@ARTICLE{intro_JPL-catalog,
       author = {{Pearson}, J.~C. and {M{\"u}ller}, H.~S.~P. and {Pickett}, H.~M. and
         {Cohen}, E.~A. and {Drouin}, B.~J.},
        title = "{Introduction to submillimeter, millimeter and microwave spectral line catalog}",
      journal = {J. Quant. Spectrosc. Radiat. Transfer},
         year = 2010,
        month = aug,
       volume = {111},
        pages = {1614-1616},
          doi = {10.1016/j.jqsrt.2010.02.002},
       adsurl = {https://ui.adsabs.harvard.edu/abs/2010JQSRT.111.1614P}
}

@ARTICLE{HCN_010_MBER_1970,
       author = {{Radford}, H.~E. and {Kurtz}, C.~V.},
        title = "{Stark effect and hyperfine structure of HCN measured with an electric resonance maser spectrometer.}",
      journal = {J. Res. Natl. Bur. Stand.},
         year = 1970,
        month = jan,
       volume = {74A},
        pages = {791-799},
       adsurl = {https://ui.adsabs.harvard.edu/abs/1970JRNBS..74..791R}
}

@ARTICLE{OCS-isos_010_MBER_1974,
       author = {{Reinartz}, J.~M.~L.~J. and {Dymanus}, A.},
        title = "{Molecular constants of OCS isotopes in the (01 $^{1}$0) vibrational state measured by molecular-beam electric-resonance spectroscopy}",
      journal = {Chem. Phys. Lett.},
         year = 1974,
        month = feb,
       volume = {24},
       number = {3},
        pages = {346-351},
          doi = {10.1016/0009-2614(74)85275-9},
       adsurl = {https://ui.adsabs.harvard.edu/abs/1974CPL....24..346R}
}

@ARTICLE{MeCN_nu4_nu7_3nu8_1993,
       author = {{Tolonen}, A.~M. and {Koivusaari}, M. and {Paso}, R. and
         {Schroderus}, J. and {Alanko}, S. and {Anttila}, R.},
        title = "{The Infrared Spectrum of Methyl Cyanide Between 850 and 1150~cm$^{-1}$: Analysis of the $\nu _{4}$, $\nu _{7}$, and $3\nu ^{1}_{8}$ Bands with Resonances}",
      journal = {J. Mol. Spectrosc.},
         year = 1993,
        month = aug,
       volume = {160},
       number = {2},
        pages = {554-565},
          doi = {10.1006/jmsp.1993.1201},
       adsurl = {https://ui.adsabs.harvard.edu/abs/1993JMoSp.160..554T}
}

@ARTICLE{pentade_1994,
       author = {{Paso}, R. and {Anttila}, R. and {Koivusaari}, M.},
        title = "{The Infrared Spectrum of Methyl Cyanide Between 1240 and 1650~cm$^{-1}$: The Coupled Band System $\nu _{3}$, $\nu _{6}^{\pm 1}$, and $(\nu _{7} + \nu _{8})^{\pm 2}$}",
      journal = {J. Mol. Spectrosc.},
         year = 1994,
        month = jun,
       volume = {165},
       number = {2},
        pages = {470-480},
          doi = {10.1006/jmsp.1994.1150},
       adsurl = {https://ui.adsabs.harvard.edu/abs/1994JMoSp.165..470P}
}

@ARTICLE{spfit_Novick_2016,
       author = {{Novick}, Stewart E.},
        title = "{A beginner's guide to Pickett's SPCAT/SPFIT}",
      journal = {J. Mol. Spectrosc.},
         year = 2016,
        month = nov,
       volume = {329},
        pages = {1-7},
          doi = {10.1016/j.jms.2016.08.015},
       adsurl = {https://ui.adsabs.harvard.edu/abs/2016JMoSp.329....1N}
}

@ARTICLE{spfit_Drouin_2017,
       author = {{Drouin}, Brian J.},
        title = "{Practical uses of SPFIT}",
      journal = {J. Mol. Spectrosc.},
         year = 2017,
        month = oct,
       volume = {340},
        pages = {1-15},
          doi = {10.1016/j.jms.2017.07.009},
       adsurl = {https://ui.adsabs.harvard.edu/abs/2017JMoSp.340....1D}
}

@misc{Fitting-Spectra,
note = {CDMS Fitting Spectra page at https://cdms.astro.uni-koeln.de/classic/pickett; accessed 2025-11-10.}
}

@ARTICLE{CDMS_2001,
       author = {{M{\"u}ller}, H.~S.~P. and {Thorwirth}, S. and {Roth}, D.~A. and {Winnewisser}, G.},
        title = "{The Cologne Database for Molecular Spectroscopy, CDMS}",
      journal = {Astron. Astrophys.},
         year = 2001,
        month = apr,
       volume = {370},
        pages = {L49-L52},
          doi = {10.1051/0004-6361:20010367},
       adsurl = {https://ui.adsabs.harvard.edu/abs/2001A&A...370L..49M}
}

@ARTICLE{CDMS_2005,
       author = {{M{\"u}ller}, Holger S.~P. and {Schl{\"o}der}, Frank and {Stutzki}, J{\"u}rgen and {Winnewisser}, Gisbert},
        title = "{The Cologne Database for Molecular Spectroscopy, CDMS: a useful tool for astronomers and spectroscopists}",
      journal = {J. Mol. Struct.},
         year = 2005,
        month = may,
       volume = {742},
       number = {1-3},
        pages = {215-227},
          doi = {10.1016/j.molstruc.2005.01.027},
       adsurl = {https://ui.adsabs.harvard.edu/abs/2005JMoSt.742..215M}
}

@ARTICLE{CDMS_2016,
       author = {{Endres}, Christian P. and {Schlemmer}, Stephan and {Schilke}, Peter and
         {Stutzki}, J{\"u}rgen and {M{\"u}ller}, Holger S.~P.},
        title = "{The Cologne Database for Molecular Spectroscopy, CDMS, in the Virtual Atomic and Molecular Data Centre, VAMDC}",
      journal = {J. Mol. Spectrosc.},
         year = "2016",
        month = "Sep",
       volume = {327},
        pages = {95-104},
          doi = {10.1016/j.jms.2016.03.005},
archivePrefix = {arXiv},
       eprint = {1603.03264},
 primaryClass = {astro-ph.IM},
       adsurl = {https://ui.adsabs.harvard.edu/abs/2016JMoSp.327...95E}
}

@ARTICLE{13C-MeCN_rot_1988,
       author = {{Tam}, H. and {An}, I. and {Roberts}, J.~A.},
        title = "{Microwave spectra of the $^{13}$C isotopic species of methyl cyanide for the ground and $\varv _{8} = 1$, 2 vibrational levels in the frequency range 17-56~GHz}",
      journal = {J. Mol. Spectrosc.},
         year = 1988,
        month = may,
       volume = {129},
       number = {1},
        pages = {202-215},
          doi = {10.1016/0022-2852(88)90270-6},
       adsurl = {https://ui.adsabs.harvard.edu/abs/1988JMoSp.129..202T}
}

@ARTICLE{MeCN-isos_rot_1996,
       author = {{Pearson}, J.~C. and {M\"uller}, H.~S.~P.},
        title = "{The Submillimeter Wave Spectrum of Isotopic Methyl Cyanide}",
      journal = {Astrophys. J.},
         year = 1996,
        month = nov,
       volume = {471},
        pages = {1067},
          doi = {10.1086/178034},
       adsurl = {https://ui.adsabs.harvard.edu/abs/1996ApJ...471.1067P}
}

@ARTICLE{MeCN-Lille_1977,
       author = {{Boucher}, D. and {Burie}, J. and {Demaison}, J. and {Dubrulle}, A. and
         {Legrand}, J. and {Segard}, B.},
        title = "{High-resolution rotational spectrum of methyl cyanide}",
      journal = {J. Mol. Spectrosc.},
         year = 1977,
        month = feb,
       volume = {64},
       number = {2},
        pages = {290-294},
          doi = {10.1016/0022-2852(77)90267-3},
       adsurl = {https://ui.adsabs.harvard.edu/abs/1977JMoSp..64..290B}
}

@ARTICLE{MeCN-12-13b_2-1,
       author = {{Kukolich}, Stephen G.},
        title = "{Beam maser spectroscopy on $J = 1 \rightarrow 2$, $K = 1$, and $K = 0$ transitions in CH$_{3}$CN and CH$_{3}$$^{13}$CN}",
      journal = {J. Chem. Phys.},
         year = 1982,
        month = jan,
       volume = {76},
       number = {1},
        pages = {97-101},
          doi = {10.1063/1.442694},
       adsurl = {https://ui.adsabs.harvard.edu/abs/1982JChPh..76...97K}
}

@ARTICLE{FF_Duncan_1978,
       author = {{Duncan}, J.~L. and {McKean}, D.~C. and {Tullini}, F. and
         {Nivellini}, G.~D. and {Perez Pe{\~n}a}, J.},
        title = "{Methyl cyanide. Spectroscopic studies of isotopically substituted species, and the harmonic potential function}",
      journal = {J. Mol. Spectrosc.},
         year = 1978,
        month = jan,
       volume = {69},
       number = {1},
        pages = {123-140},
          doi = {10.1016/0022-2852(78)90033-4},
       adsurl = {https://ui.adsabs.harvard.edu/abs/1978JMoSp..69..123D}
}

@ARTICLE{HCN_rot_v2le1_2003,
       author = {{Thorwirth}, S. and {M{\"u}ller}, H.~S.~P. and {Lewen}, F. and {Br{\"u}nken}, S. and {Ahrens}, V. and {Winnewisser}, G.},
        title = "{A Concise New Look at the $l$-Type Spectrum of H$^{12}$C$^{14}$N}",
      journal = {Astrophys. J.},
         year = 2003,
        month = mar,
       volume = {585},
       number = {2},
        pages = {L163-L165},
          doi = {10.1086/374327},
       adsurl = {https://ui.adsabs.harvard.edu/abs/2003ApJ...585L.163T}
}

@ARTICLE{MeCCH_nu10+Dyade_2002,
       author = {{M{\"u}ller}, H.~S.~P. and {Pracna}, P. and {Horneman}, V. -M.},
        title = "{The $\varv _{10}=1$ Level of Propyne, H$_{3}$C$-$C$\equiv$CH, and Its Interactions with $\varv _{9}=1$ and $\varv _{10}=2$}",
      journal = {J. Mol. Spectrosc.},
         year = 2002,
        month = dec,
       volume = {216},
       number = {2},
        pages = {397-407},
          doi = {10.1006/jmsp.2002.8661},
       adsurl = {https://ui.adsabs.harvard.edu/abs/2002JMoSp.216..397M}
}

@ARTICLE{CH3CCH_v39_2004,
       author = {{Pracna}, P. and {M{\"u}ller}, H.~S.~P. and {Klee}, S. and
         {Horneman}, V. -M.},
        title = "{Interactions in symmetric top molecules between vibrational polyads: rotational and rovibrational spectroscopy of low-lying states of propyne, H$_3$C$-$C$\equiv$CH}",
      journal = {Mol. Phys.},
         year = "2004",
        month = "Jan",
       volume = {102},
       number = {14},
        pages = {1555-1568},
          doi = {10.1080/00268970410001725864},
       adsurl = {https://ui.adsabs.harvard.edu/abs/2004MolPh.102.1555P}
}

@ARTICLE{MeCCH_10mue_2009,
       author = {{Pracna}, P. and {M{\"u}ller}, H.~S.~P. and {Urban}, {\v{S}}. and
         {Horneman}, V. -M. and {Klee}, S.},
        title = "{Interactions between vibrational polyads of propyne, H$_{3}$C$-$C$\equiv$CH: Rotational and rovibrational spectroscopy of the levels around 1000~cm$^{-1}$}",
      journal = {J. Mol. Spectrosc.},
         year = "2009",
        month = "Jul",
       volume = {256},
       number = {1},
        pages = {152-162},
          doi = {10.1016/j.jms.2009.04.003},
       adsurl = {https://ui.adsabs.harvard.edu/abs/2009JMoSp.256..152P}
}

@ARTICLE{MeNC_v8le2_2011,
       author = {{Pracna}, P. and {Urban}, J. and {Votava}, O. and {Meltzerov{\'a}}, Z. and
         {Urban}, {\v{S}}. and {Horneman}, V. -M.},
        title = "{Rotational and rovibrational spectroscopy of the $\varv_{8} = 1$ and 2 vibrational states of CH$_{3}$NC}",
      journal = {Mol. Phys.},
         year = "2011",
        month = "Sep",
       volume = {109},
       number = {17-18},
        pages = {2237-2243},
          doi = {10.1080/00268976.2011.605775},
       adsurl = {https://ui.adsabs.harvard.edu/abs/2011MolPh.109.2237P}
}

@ARTICLE{Weeds_2011,
       author = {{Maret}, S. and {Hily-Blant}, P. and {Pety}, J. and {Bardeau}, S. and {Reynier}, E.},
        title = "{Weeds: a CLASS extension for the analysis of millimeter and sub-millimeter spectral surveys}",
      journal = {Astron. Astrophys.},
         year = 2011,
        month = feb,
       volume = {526},
          eid = {A47},
        pages = {A47},
          doi = {10.1051/0004-6361/201015487},
archivePrefix = {arXiv},
       eprint = {1012.1747},
 primaryClass = {astro-ph.IM},
       adsurl = {https://ui.adsabs.harvard.edu/abs/2011A&A...526A..47M}
}

@ARTICLE{13C-VyCN_2008,
       author = {{M{\"u}ller}, Holger S.~P. and {Belloche}, Arnaud and {Menten}, Karl M. and {Comito}, Claudia and {Schilke}, Peter},
        title = "{Rotational spectroscopy of isotopic vinyl cyanide, H$_{2}$CCHCN, in the laboratory and in space}",
      journal = {J. Mol. Spectrosc.},
         year = 2008,
        month = sep,
       volume = {251},
       number = {1-2},
        pages = {319-325},
          doi = {10.1016/j.jms.2008.03.016},
archivePrefix = {arXiv},
       eprint = {0806.2098},
 primaryClass = {astro-ph},
       adsurl = {https://ui.adsabs.harvard.edu/abs/2008JMoSp.251..319M}
}

@ARTICLE{ROH_RSH_EMoCA_2016,
       author = {{M{\"u}ller}, Holger S.~P. and {Belloche}, Arnaud and {Xu}, Li-Hong and {Lees}, Ronald M. and {Garrod}, Robin T. and {Walters}, Adam and {van Wijngaarden}, Jennifer and {Lewen}, Frank and {Schlemmer}, Stephan and {Menten}, Karl M.},
        title = "{Exploring molecular complexity with ALMA (EMoCA): Alkanethiols and alkanols in Sagittarius B2(N2)}",
      journal = {Astron. Astrophys.},
         year = 2016,
        month = mar,
       volume = {587},
          eid = {A92},
        pages = {A92},
          doi = {10.1051/0004-6361/201527470},
archivePrefix = {arXiv},
       eprint = {1512.05301},
 primaryClass = {astro-ph.GA},
       adsurl = {https://ui.adsabs.harvard.edu/abs/2016A&A...587A..92M}
}

@ARTICLE{12C-13C_SgrB2N_2017,
       author = {{Halfen}, D.~T. and {Woolf}, N.~J. and {Ziurys}, L.~M.},
        title = "{The $^{12}$C/$^{13}$C Ratio in Sgr B2(N): Constraints for Galactic Chemical Evolution and Isotopic Chemistry}",
      journal = {Astrophys. J.},
         year = "2017",
        month = "Aug",
       volume = {845},
       number = {2},
          eid = {158},
        pages = {158},
          doi = {10.3847/1538-4357/aa816b},
       adsurl = {https://ui.adsabs.harvard.edu/abs/2017ApJ...845..158H}
}

@ARTICLE{det-13CH_2020,
       author = {{Jacob}, Arshia M. and {Menten}, Karl M. and {Wiesemeyer}, Helmut and {G{\"u}sten}, Rolf and {Wyrowski}, Friedrich and {Klein}, Bernd},
        title = "{First detection of $^{13}$CH in the interstellar medium}",
      journal = {Astron. Astrophys.},
         year = 2020,
        month = aug,
       volume = {640},
          eid = {A125},
        pages = {A125},
          doi = {10.1051/0004-6361/201937385},
archivePrefix = {arXiv},
       eprint = {2007.01190},
 primaryClass = {astro-ph.GA},
       adsurl = {https://ui.adsabs.harvard.edu/abs/2020A&A...640A.125J}
}

@misc{CDMS-data,
note = {The present CH$_3$CN data are available at https://cdms.astro.uni-koeln.de/classic/predictions/daten/CH3CN/ in different subfolders; accessed 2026-01-08.}
}

@misc{CDMS-entries,
note = {See https://cdms.astro.uni-koeln.de/classic/entries/; accessed 2026-01-08.}
}

\end{document}